\documentclass[a4paper,fleqn,usenatbib]{mnras}

\usepackage{times}

\usepackage[T1]{fontenc}
\usepackage{ae,aecompl}
\usepackage{pdflscape}
\usepackage{fdsymbol}

\usepackage{graphicx}	
\usepackage{amsmath}	
\usepackage{color}
\usepackage[dvipsnames]{xcolor}
\usepackage{hyperref}
\newcommand{\xmm}{\textit{XMM-Newton} }
\newcommand{\ns}{\textit{NuSTAR} }
\newcommand{\source}{J1706}

\defcitealias{degenaar17}{D17} 
\defcitealias{Strohmayer17}{SK17}
\defcitealias{vandeneijnden17}{vdEijnden17}

\title[The very faint nature of IGR J17062--6143]{The very-faint X-ray binary IGR J17062-6143:\\ a truncated disk, no pulsations and a possible outflow}

\author[Van den Eijnden et al.]{
\noindent J. van den Eijnden$^{1,2}$\thanks{E-mail: a.j.vandeneijnden@uva.nl},
N. Degenaar$^{1,2}$, 
C. Pinto$^{2}$,
A. Patruno$^{3,4}$,
K. Wette$^{5}$, 
 \newauthor C. Messenger$^{6}$,
J. V. Hern\'andez Santisteban$^{1,2}$,
R. Wijnands$^{1}$,
J. M. Miller$^{7}$,
 \newauthor D. Altamirano$^{8}$, F. Paerels$^{9,10}$, D. Chakrabarty$^{11}$ and A. C. Fabian$^{2}$
\\
$^{1}$Anton Pannekoek Institute for Astronomy, University of Amsterdam, Science Park 904, 1098 XH Amsterdam, The Netherlands\\
$^{2}$Institute of Astronomy, University of Cambridge, Madingley Road, Cambridge CB3 0HA, UK\\
$^{3}$Leiden Observatory, Leiden University, P.O. Box 9513, NL-2300 RA Leiden, The Netherlands\\
$^{4}$ASTRON, The Netherlands Institute for Radio Astronomy, Postbus 2, 7990 AA Dwingeloo, The Netherlands\\
$^{5}$Albert-Einstein-Institut, Max-Planck-Institut f\"ur Gravitationsphysik, D-30167 Hannover, Germany\\
$^{6}$SUPA, School of Physics and Astronomy, University of Glasgow, Glasgow G12 8QQ, UK\\
$^{7}$Department of Astronomy, University of Michigan, 500 Church Street, Ann Arbor, MI 48109, USA\\
$^{8}$Department of Physics and Astronomy, University of Southampton, Southampton, Hampshire, SO171BJ, UK\\
$^{9}$Columbia University, Mail Code 5246, 550 West 120th Street, New York, NY 10027, USA\\
$^{10}$Columbia Astrophysics Laboratory, Mail Code 5247, 550 West 120th Street, New York, NY 10027, USA\\
$^{11}$Massachusetts Institute of Technology (MIT), Kavli Institute for Astrophysics and Space Research, Cambridge, MA 02139, USA\\
}

\date{Accepted XXX. Received YYY; in original form ZZZ}

\pubyear{2017}

\begin{document}
\label{firstpage}
\pagerange{\pageref{firstpage}--\pageref{lastpage}}
\maketitle

\begin{abstract}

\noindent We present a comprehensive X-ray study of the neutron star low-mass X-ray binary IGR J17062-6143, which has been accreting at low luminosities since its discovery in $2006$. Analysing \textit{NuSTAR}, \textit{XMM-Newton} and \textit{Swift} observations, we investigate the very faint nature of this source through three approaches: modelling the relativistic reflection spectrum to constrain the accretion geometry, performing high-resolution X-ray spectroscopy to search for an outflow, and searching for the recently reported millisecond X-ray pulsations. We find a strongly truncated accretion disk at $77^{+22}_{-18}$ gravitational radii ($\sim 164$ km) assuming a high inclination, although a low inclination and a disk extending to the neutron star cannot be excluded. The high-resolution spectroscopy reveals evidence for oxygen-rich circumbinary material, possibly resulting from a blueshifted, collisionally-ionised outflow. Finally, we do not detect any pulsations. We discuss these results in the broader context of possible explanations for the persistent faint nature of weakly accreting neutron stars. The results are consistent with both an ultra-compact binary orbit and a magnetically truncated accretion flow, although both cannot be unambigiously inferred. We also discuss the nature of the donor star and conclude that it is likely a CO or O-Ne-Mg white dwarf, consistent with recent multi-wavelength modelling.
\end{abstract}




\section{Introduction}

In low-mass X-ray binaries (LMXBs), either a neutron star (NS) or a black hole (BH) accretes matter from a low-mass companion star overflowing its Roche lobe. Such LMXBs typically are transient systems, displaying outbursts lasting weeks to months and afterwards returning to quiescence for months to years. Around the peak of these outbursts, where the accretion rate typically reaches few tens of percents of the Eddington rate, the accretion flow is well described by a geometrically thin, optically thick accretion disk extending to the compact object. At lower accretion rates, an additonal, poorly-understood Comptonizing structure of hot electrons, the corona, is typically located close to the compact object \citep[see e.g.][for reviews]{done07,gilfanov10}. At such lower accretion rates, the accretion flow also changes its structure significantly \citep{wagner94,campana97,rutledge02,kuulkers09,cackett13,bernardini13,chakrabarty14,dangelo15,rana16}: the inner flow is predicted to transition into a radiatively ineffecient accretion flow (RIAF) as the thin disk evaporates into a hot, thick flow \citep{narayan94,blandford99,menou00,dubus01}. Understanding this low-level regime of accretion is essential, albeit challenging, to form a complete picture of accretion physics in LMXBs. 

In both NS and BH LMXBs, the spectrum is observed to become softer as the source becomes fainter, possibly indicating the formation of a hot, thick inner flow \citep{armas11, armas13b, degenaar13, bahramian14, allen15, weng15}. However, the presence of a possible solid surface and an anchored magnetic field in NSs creates several differences compared to BHs at these lower accretion rates. Firstly, when high-quality spectral data is available at lower rates, a thermal component unobserved in BHs emerges in NSs; this component is thought to originate from the accretion-heated NS surface \citep{wijnands15}. Additionally, a 
harder power-law tail is observed in NSs than in BHs \citep{armas13a, armas13b, degenaar13, wijnands15}. Finally, the magnetic field of the NS can interact with the accretion flow, possibly truncating the disk away from the compact object \citep[e.g.][]{illarionov75, cackett10, dangelo10}. As the gas pressure decreases towards lower accretion rates, this interaction and truncation might be more efficient in this accretion regime. Disk truncations have been inferred in a few NS LMXBs at larger radii than in BHs at similar accretion rates, possible indeed caused by the NS magnetic fields \citep[e.g.][see Appendix A for a detailed comparison.]{tomsick09, degenaar14, furst16, iaria16, degenaar17,  vandeneijnden17, ludlam17b}.

The low-luminosity epochs during the outbursts decays in transient LMXBs are challenging to study due to the short timescales and low fluxes involved. However, interestingly, a small sample of NS LMXBs is observed to accrete in this transition regime persistently for years \citep[$L_X \sim 10^{-4}$--$10^{-2} L_{\rm Edd}$, where $L_{\rm Edd}$ is the Eddington luminosity, corresponding to the maximum possible accretion rate][]{chelovekov07, delsanto07, jonker08, heinke09, intzand09, degenaar10, degenaar17, armas13a}. These sources, called very-faint X-ray binaries or VFXBs, are thus interesting to study the low-level accretion regime in between outburst and quiescence. However, these sources evidently have an additional complication: it is currently unclear how they can persistently accrete at such low levels, and this persistent nature might make their faint properties different from transient sources. 

Two different explanations have been proposed to account for the persistently faint nature of VFXBs: magnetic inhibition of the accretion flow and an ultra-compact nature of the binary. In the former, a strong NS magnetic field truncates the inner accretion disk, effectively preventing efficient accretion \citep{ heinke15, degenaar14, degenaar17}. In this scenario, the field lines might act as a magnetic propeller, which could cause the expulsion of gas into an outflow and reduces the accretion efficiency. Alternatively, only a small accretion disk physically fits into the compact binary orbit of a so-called ultra-compact X-ray binary, or UCXB \citep{king06, intzand07, hameury16}. This second scenario can evidently be tested directly by measuring the orbital parameters. More indirectly, as the small orbit does not fit a hydrogen-rich donor \citep[e.g.][]{nelemans04, intzand09}, a lack of hydrogen emission from the accretion disk can hint towards a ultra-compact orbit. However, several LMXBs lacking hydrogen emission without having an ultra-compact orbit have been detected and additionally, a VFXB with hydrogen emission has also been observed \citep{degenaar10}. Furthermore, these two mechanisms are not necessarily conflicting: a strong magnetic field NS can be located in a UCXB system, as in for instance 4U 1626-67 \citep{chakrabarty97}. 

In this paper, we investigate the VFXB IGR~J17062--6143 (hereafter J1706). Discovered in 2006 by the \textit{INTEGRAL} satellite \citep{churazov07}, it has persistently hovered between a luminosity of $\sim 10^{-3}$--$10^{-2}$ $L_{\rm Edd}$, for the past decade. Since its discovery, it has not been observed to either go into outburst or return to quiescence. Its neutron star nature was identified by the detection of a Type-I X-ray burst in 2012 \citep{degenaar13}, which also yielded a distance estimate of $\sim 5$ kpc. More recently, the analysis of a second Type-I burst in 2015 resulted in a larger estimated distance of $7.3 \pm 0.5$ kpc \citep{keek17}. In this work, we adopt this second, more recent, and likely more accurate distance estimate. 

The X-ray spectral properties of \source~were studied by \citet[hereafter D17]{degenaar17}, analysing simultaneous \textit{Swift}, \textit{Chandra} and \textit{NuSTAR} observations. The \textit{NuSTAR} and \textit{Chandra} spectra clearly revealed a broad iron-K line around $6.5$ keV, for the first time at such a low ($2.5 \times 10^{-3} L_{\rm Edd}$) accretion rate in an NS LMXB. This iron-K line is the most prominent feature of the reflection spectrum: photons originating from close to the compact object (for instance from the Comptonizing hot flow) reflecting off the disk into our line of sight. The iron-K line profile feature is altered into a broadened shape by the rotation of the disk, gravitational redshift and relativistic boosting \citep{fabian89}. Hence, by modelling both this line and the remainder of the reflection spectrum, it is possible to infer geometrical parameters such as the inner disk radius and inclination of the system. 

Through detailed modelling of the reflection spectrum, \citetalias{degenaar17} inferred that the accretion disk is truncated far from the NS at $R_{\rm in} \gtrsim 100$ $R_g$, where $R_g = GM/c^2$ is the gravititional radius ($\sim 2.07$ km for a $1.4$ $M_{\odot}$ NS). Although the innermost stable circular orbit (ISCO; $6$ $R_g$ for a non-spinning compact object) could not be excluded at $3\sigma$, this inferred inner radius is significantly larger than typically observed in accreting neutron stars. At these low accretion rates, it is difficult to definitively distinguish between the NS's magnetic field truncating the disk, or the formation of a hot inner flow resulting in a large inner disk radius. However, for \source, the inferred inner radius is also significantly larger than observed in two BH LMXBs at similar or lower accretion rates: $\geq 35$ $R_g$ in GX 339-4 \citep{tomsick09} and $12-35$ $R_g$ in GRS 1739-278 \citep{furst16}. As the formation of a hot flow might also be more efficient in BHs, due to the lack of photons from the NS surface cooling the flow \citep[e.g.][]{narayan95}, \citetalias{degenaar17} concluded that the disk in \source~is likely truncated by the magnetic field. Under that assumption, the measured flux and $R_{\rm in}$ predict a magnetic field of $B \gtrsim 4\times 10^8$ G.  

Additionally, \citetalias{degenaar17} performed high-resolution X-ray spectroscopy on the \textit{Chandra}-HETG spectra. Several (marginally) significant emission and absorption lines could be detected, although an unambiguous identification was not possible. The presence of blueshifted absorption suggests the presence of a wind, which might be driven by a propeller resulting from the magnetic truncation of the disk or alternatively radiation pressure in the disk. Interestingly, if the outflow is propellor-driven, combined with the possible magnetic truncation of the accretion disk, this appears to be consistent with the idea of magnetic inhibition in VFXBs introduced above. However, due to the low flux of \source, both results are merely marginally significant and require independent confirmation with new observations. The recent detection of $163$ Hz coherent X-ray pulsations in J1706 by \citet{Strohmayer17} is consistent with this picture of a magnetically truncated disk. 

However, evidence for an ultracompact nature of \source~was also recently found. {\color{blue} Hern\'andez Santisteban et al. (2017)} performed a multi-wavelength study covering the optical, UV and NIR. Optical \textit{Gemini} spectroscopy revealed a blue but featureless disk spectrum, consistent with a hydrogen-poor donor star, as is expected in UCXBs \citep{intzand09}. In addition, the modelling of the complete disk spectral-energy distribution (SED) provides an estimate of the orbital period of $0.6$--$1.3$ hour. Hence, arguments can be made both for an ultracompact nature and for magnetic inhibition of the accretion flow in \source, and new, detailed observational studies are required to fully understand its persistently low accretion rate.

In this paper, we present a detailed study of new and archival X-ray observations of \source~by \textit{NuSTAR}, \textit{XMM-Newton} and \textit{Swift}, aiming to understand its VFXB nature through three approaches: high-resolution X-ray spectroscopy of the \textit{XMM-Newton} RGS spectra, broadband reflection modelling of all observations, and finally an extensive pulsation search in the \textit{XMM-Newton} EPIC-pn data. While the low flux of VFXBs makes each of these individual methods challenging, their combination yields firmer constraints on the accretion properties of J1706. 

\section{Observations}
\label{sec:observations}

We extended the set of observations of \source~analysed by \citetalias{degenaar17}, which consisted of \textit{NuSTAR}, \textit{Swift} (both from 2015) and \textit{Chandra} (from 2014) observations, with new, simultaneous \ns and \xmm observations from September 2016. For the 2015 observations, we applied the same approach to the data reduction as \citetalias{degenaar17}. For clarity, we briefly review that approach in this section, in addition to a more detailed discussion of the 2016 data. During the 2016 observations, \source~shows a $\sim 16\%$ lower luminosity than in the 2015 data; we will discuss the similarities and discrepancies between the two datasets in Section 3. We included the 2015 \textit{Swift} observation to increase the soft spectral coverage during the 2015 epoch. We did not reanalyse the \textit{Chandra}-observation, but instead focused on \xmm RGS in our search for narrow line features. During none of the analysed observation a Type-I burst was observed. 

\subsection{\ns} 

\ns \citep{harrison13} observed \source~from 19:26:07 May 6 to 05:01:07 May 8 2015 (ObsId 30101034002) and from 08:46:08 September 13 to 14:36:08 September 14 2016 (ObsId 30101018002). We applied the standard \textsc{nupipeline} and \textsc{nuproducts} software to extract source and background spectra, and lightcurves, for both observations. The 2015 and 2016 observations amount to a $\sim 70$ and $\sim 67$ ks exposure, respectively. Following \citetalias{degenaar17}, we selected a $30$ arcsec circular source region and a $60$ arcsec background region from the same chip in both observations. As in \citetalias{degenaar17}, we found a neglegible ($<0.5\%$) difference in normalisation between the Focal Plane Module A and B (FPMA/FPMB) spectra in the 2015 observation - hence, we combined the data from the two modules using \textsc{addascaspec} and \textsc{addrmf}. The 2016 obseration shows larger deviations between FPMA and FPMB ($\sim 6\%$), and are thus not combined but rather fitted simultaneously with a constant floating in between. Finally, we rebinned the combined 2015 spectrum and two seperate 2016 FPMA and FPMB spectra to contain at least 20 counts per bin. \source~is detected above the background in the entire $3$--$79$ keV bandpass in the 2015 observation, and in the $3$--$50$ keV range in the 2016 data. 

\subsection{\textit{Swift}}

The \textit{Swift} \citep{gehrels04} X-ray Telescope (XRT) observed \source~in Photon Counting mode on May 6 2015 (ObsId 00037808005), simultaneously with the first \ns observation, amounting to a $\sim 0.9$ ks exposure. We again followed the extraction approach in \citetalias{degenaar17}. Using \textsc{xselect}, we extracted a source spectrum from a $12$--$71$ arcsec annulus to circumvent pile-up issues, and a background spectrum from a void region three times the size. We produced an arf file with \textsc{xrtmkarf} and used the appropriate rmf file (version 15: \textsc{swxwt0to2s6\_20131212v015.rmf}) from the \textsc{caldb}. Finally, we rebinned the spectrum to contain a minimum of 20 counts per bin. 

\subsection{\xmm}

\xmm \citep{jansen01} observed \source~from 12:21:18 September 13 to 06:04:51 September 15 2016 (ObsId 0790780101). We extracted spectra from the EPIC-pn, which operated in timing mode, and RGS detectors using the \xmm \textsc{SAS} v15 following the standard procedures in the \textsc{SAS} cookbook\footnote{https://heasarc.gsfc.nasa.gov/docs/xmm/abc/}. The EPIC-pn $10$--$12$ keV lightcurve does not show any background flaring, so we used all available data. We extracted the EPIC-pn source and background spectra from regions of RAWX between $30$ and $46$, and between $2$ and $6$, respectively. Using the \textsc{ftool} \textsc{epatplot}, we explicitly checked for pile-up in the spectrum, which is not present. We extracted the RGS spectra following the standard \textsc{SAS} guidelines, combining the two detectors into one spectrum per order after visually confirming that the two detectors are consistent. We analyse the resulting RGS first and second order spectra in the $7.0$--$28.0$ \AA~and $7.0$--$16.0$ \AA~wavelength ranges, respectively. 

\section{Broadband spectral analysis}
\label{sec:spec}

We fitted the X-ray spectra using \textsc{xspec} v12.9.0 \citep{arnaud96}. In order to model the interstellar absorption, we included either \textsc{tbabs} or \textsc{tbnew} in each model, depending on whether we use Solar abundances in the absorbtion column. We used cross-sections from \citet{verner96} and Solar abundances from \citet{wilms00}. In addition, we inluded a floating constant between all spectra to account for normalisation offsets between the data sets. We quote uncertainties at the $1\sigma$ level. 

\subsection{Phenomenological modelling}
\label{sec:phenom}

\citetalias{degenaar17} phenomenologically described the 2015 \textit{Swift} and \ns spectra with a model consisting of a powerlaw (\textsc{pegpwrlw}) and a blackbody (\textsc{bbodyrad}). To investigate the similarity between the spectra from the 2015 and 2016 observations, we first applied the same model to the 2016 data only -- note that we did not include the RGS spectrum in this broadband modelling, but instead seperately focus on it in Section \ref{sec:highres}. Due to the increased quality of the \xmm EPIC-pn spectrum compared to the \textit{Swift} spectrum, this phenomenological model does not provide an adequate description below $3$ keV ($\chi^2_{\nu} \sim 2.9$ for $849$ degrees of freedom). Large residuals remain around $1$ keV, which cannot be described by an additional \textsc{diskbb} component representing the accretion disk ($\chi^2_{\nu} \sim 2.7$). Instead including an additional Gaussian component at this energy resulted in a highly improved fit ($\chi^2_{\nu} = 1083.8/846 = 1.28$), with the Gaussian centroid energy and width equal to $0.96 \pm 0.01$~keV and $0.18 \pm 0.01$~keV, respectively (see Section \ref{sec:highres} for a detailed analysis of this feature). The power law index equals $\Gamma = 2.00 \pm 0.01$, while the blackbody temperature and radius are $T_{\rm BB} = 0.365 \pm 0.003$~keV and $R_{\rm BB} = (6.96 \pm 0.2) [D/7.3 \rm kpc]$~km, respectively. Interestingly, this blackbody temperature is significantly lower than the $T_{\rm BB} = 0.46 \pm 0.03$~keV found by \citetalias{degenaar17}. 

\begin{figure}
  \begin{center}
    \includegraphics[width=\columnwidth]{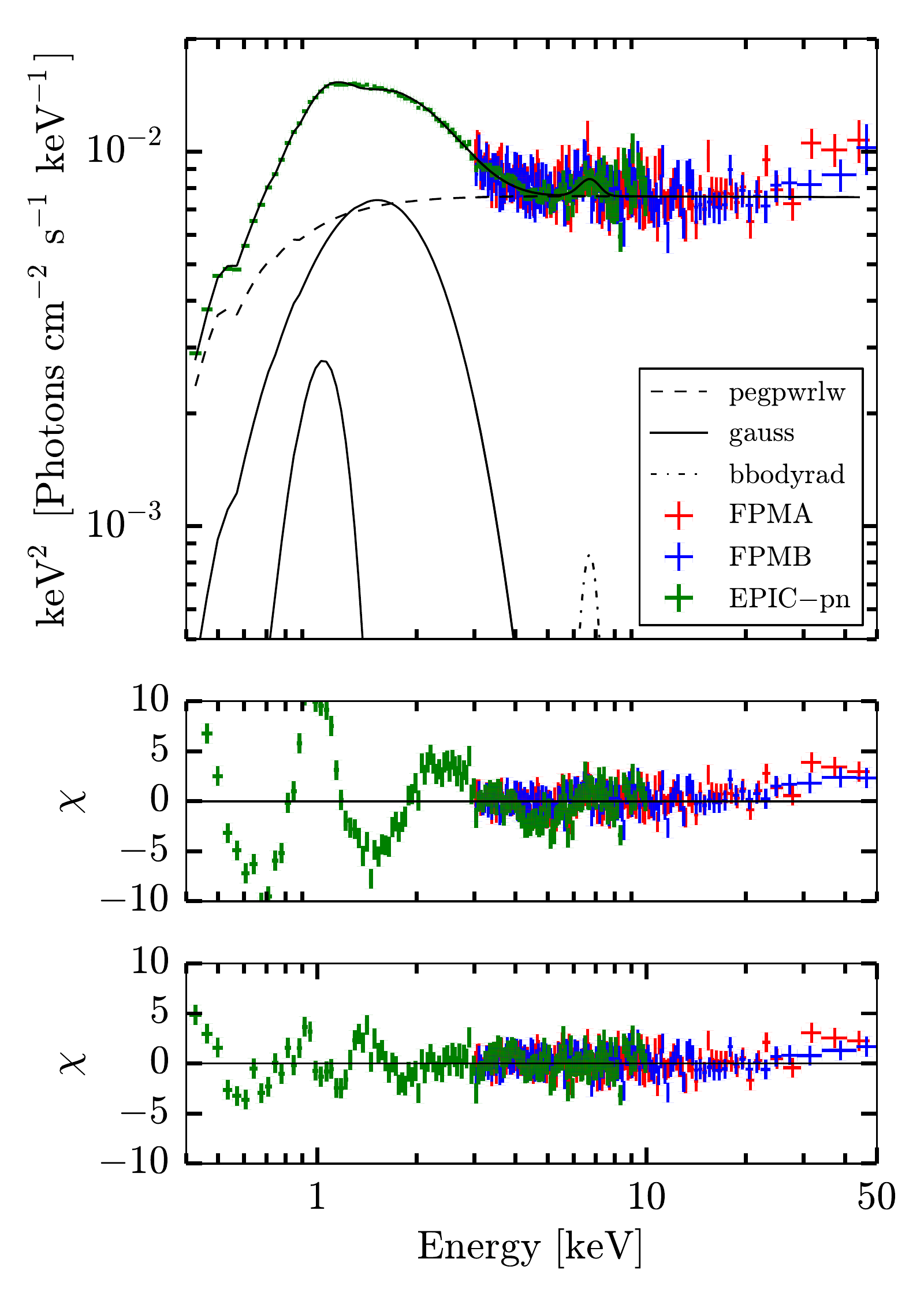}
    \caption{\textit{Top:} The 2016 \textit{XMM-Newton} EPIC-pn (green), and \textit{NuSTAR} FPMA (red) and FPMB (blue) spectra unfolded around the best-fitting phenomenological model \textsc{constant*tbabs*(bbodyrad+pegpwrlw+gauss+gauss)}. The FPMB and EPIC-pn spectra have been rescaled by their fitted cross-calibration constant for visual clarity. \textit{Middle:} $\chi$ for the phenomenological model in \citetalias{degenaar17}, showing a broad Fe K$\alpha$ feature and an emission feature around $\sim 1$~keV. \textit{Bottom:} $\chi$ for the best fitting phenomenological model. We discuss the residuals remaining below $2$ keV in detail in Section \ref{sec:highres}.}
    \label{fig:spec1}
  \end{center}
\end{figure}

\begin{figure}
  \begin{center}
    \includegraphics[width=\columnwidth]{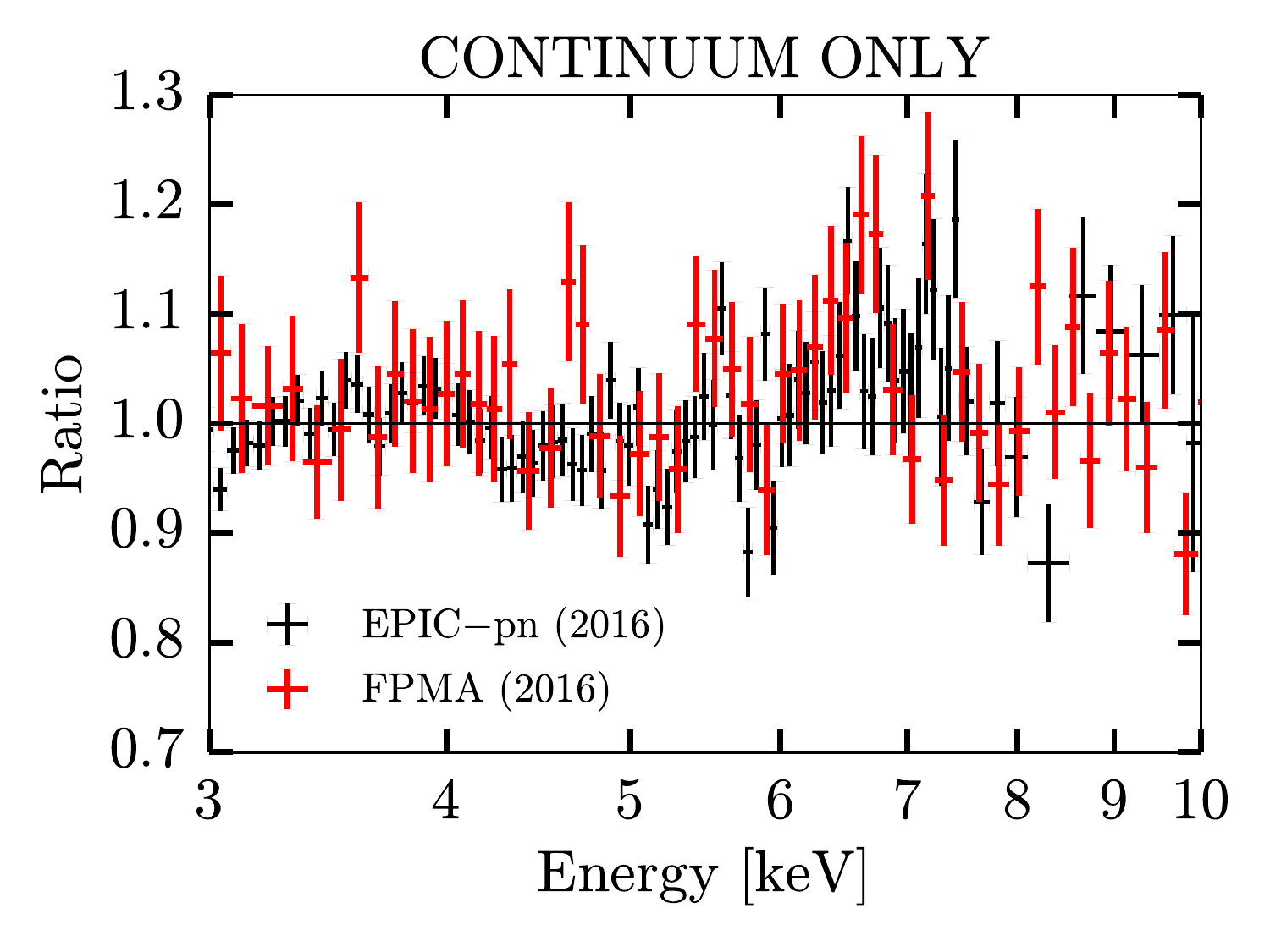}
    \caption{Data-to-model ratio of the 2016 \textit{XMM-Newton} EPIC-pn (black) and \textit{NuSTAR} FPMA (red) spectra fitted with a simple continuum model (see Section \ref{sec:phenom}). An excess emission feature around the Fe K$\alpha$ energy ($6.4$--$6.97$ keV) is clearly visible. The spectrum has been rebinned for visual purposes.}
    \label{fig:spec4}
  \end{center}
\end{figure}

In Figure \ref{fig:spec1}, we show the unfolded 2016 spectra in the upper panel, and the residuals for the \citetalias{degenaar17}-phenomenological model in the middle panel. The 2015 observations of \source~contain a significant Fe K$\alpha$ line; zooming in on the residuals of the 2016 data fitted to the continuum model (see Figure \ref{fig:spec4}), suggests the presence of a similar broad feature. To test whether this broad line is significant in the 2016 data as well, we added a Gaussian line to the phenomenological model with a centroid energy constrained to $6.4$--$6.97$~keV (the possible range for Fe K$\alpha$ emission). This results in a better fit, with $\chi^2_{\nu} = 1020.32/843 = 1.21$ (f-test rejection probability of $5\times10^{-11}$) and a line normalisation of $(2.2\pm0.35)\times10^{-5}$ photons cm$^{-2}$s$^{-1}$. The resulting Gaussian parameters are a centroid energy of $6.65\pm 0.08$~keV, a width of $0.47^{+0.09}_{-0.08}$~keV, and an equivalent width of $\rm EW = 120$~eV, all within the typical range for Gaussian iron lines \citep[e.g.][]{ng10}. 

The bottom panel in Figure \ref{fig:spec1} shows the residuals after the inclusion of the two Gaussian features at $\sim 1.0$ keV and $\sim 6.66$~keV. Some residuals below $2$ keV remain after the inclusion of a Gaussian around $1$ keV; we will investigate the nature of these residuals in detail in Section \ref{sec:highres}. Finally, slightly positive residuals are present above $\sim 30$ keV. However, the inclusion of a second powerlaw component \citep[as in][]{degenaar17} does not significantly improve the fit ($p=0.01$ for an unphysical power-law index of $\Gamma = -2.5$). Refitting the continuum model up to $30$ keV only does not result in any changes in the parameters, so these residuals do not influence the fit. We also note that an absorption feature appears to be present at $\sim 8.2$~keV. However, as this feature is only present in the EPIC-pn spectrum (see e.g. Figure \ref{fig:spec4}), it most probably originates from known Ni, Cu and Zn fluorescence lines in the internal instrument background spectrum around this energy.\footnote{See section 3.3.7.2 in the XMM-Newton Users Handbook}  

Extending the best-fitting phenomenological model to the $0.3$--$79$ keV range yields an unabsorbed flux of $(0.98\pm0.02)\times10^{-10}$ erg s$^{-1}$cm$^{-2}$, which is only slightly lower than the flux during the 2015 observations ($(1.17\pm0.02)\times10^{-10}$ erg s$^{-1}$cm$^{-2}$). Given this similarity in flux, spectral shape and parameters (apart from the $\sim 1$~keV excess), and the presence of a Fe K$\alpha$ line, we subsequently fitted the 2015 and 2016 observations together. 

\subsection{Relativistic reflection models}
\label{sec:relref}

\subsubsection{The iron line: \textsc{diskline}}
Before including relativistic reflection in our spectral model, we first analysed the continuum in the 2015 and 2016 observations together. Simply applying a model consisting of \textsc{pegpwrlw}, \textsc{bbodyrad} and a Gaussian around $1$~keV with all parameters tied results in a bad fit, with $\chi^2_{\nu} = 1905.3/1436 = 1.33$. This is not surprising given the difference in flux, and so we check which parameters differ significantly between the two epochs. Inspection of the residuals reveals clear differences between the two datasets below $3$~keV and a possibly different powerlaw index. Indeed, untying the blackbody temperature and radius results in a significantly improved fit ($\chi^2_{\nu} = 1730.0/1434 = 1.21$; f-test rejection probability $p \sim 10^{-31}$). In addition, untying the powerlaw index also results in a marginally significant improvement ($\chi^2_{\nu} = 1705.4/1433 = 1.19$; $p=6\times10^{-6}$), with a slightly harder spectrum in 2016 ($\Gamma = 2.00 \pm 0.01$ compared to $2.08 \pm 0.01$). Untying the powerlaw normalisation however does not result in a significant improvement of the fit, both when the powerlaw index is tied between the two epochs or free. All parameters of the final continuum model are listed in Table \ref{tab:pars}. 

We first modeled the Fe K$\alpha$ line using the \textsc{diskline}-model \citep{fabian89}, which models a single emission line from the accretion disk, assuming a Schwarzschild metric, e.g. a dimensionless spin parameter of $a=0.0$. For NSs, the spin $a$ typically ranges from $0.0$ to $0.3$, where it only minimally impacts the surrounding metric. Initially, we do not link the \textsc{diskline} parameters between the 2015 and 2016 observations. As in the 2015 observations alone, the inclination is ill-constrained in the 2016 observations. As $\chi^2$ is minimum at $i\approx 67-69^{\rm o}$, we followed \citetalias{degenaar17} and initially fixed the inclination to $65^{\rm o}$. The fitted iron line parameters (line energy, inner disk radius and normalisation) are all consistent between the 2015 and 2016 epochs. Hence, to increase the accuracy of our parameter determination, we link all three between the 2015 and 2016 spectra. The resulting fit ($\chi^2_{\nu} = 1580.8/1430 = 1.11)$ implies an inner disk radius of $R_{\rm in} = 77^{+22}_{-18}$ $R_g$, with the ISCO excluded at a significance of $\sim 6.2 \sigma$. 

As we will discuss in Section \ref{sec:bb} in detail, the reflecting disk itself is not observed in the X-rays. It is however clearly observed in the source's SED ({\color{blue} Hern\'andez Santisteban et al., 2017.}). The inner disk radius measurement from reflection spectroscopy is consistent with the modelling of this SED, as the SED only constrains the inner radius to be larger than the NS radius. Detecting the accretion disk in the SED but not in the X-ray spectrum is consistent with a large truncation radius, as the accretion disk X-ray emission originates from the innermost regions.

\begin{figure*}
  \begin{center}
    \includegraphics[width=\textwidth]{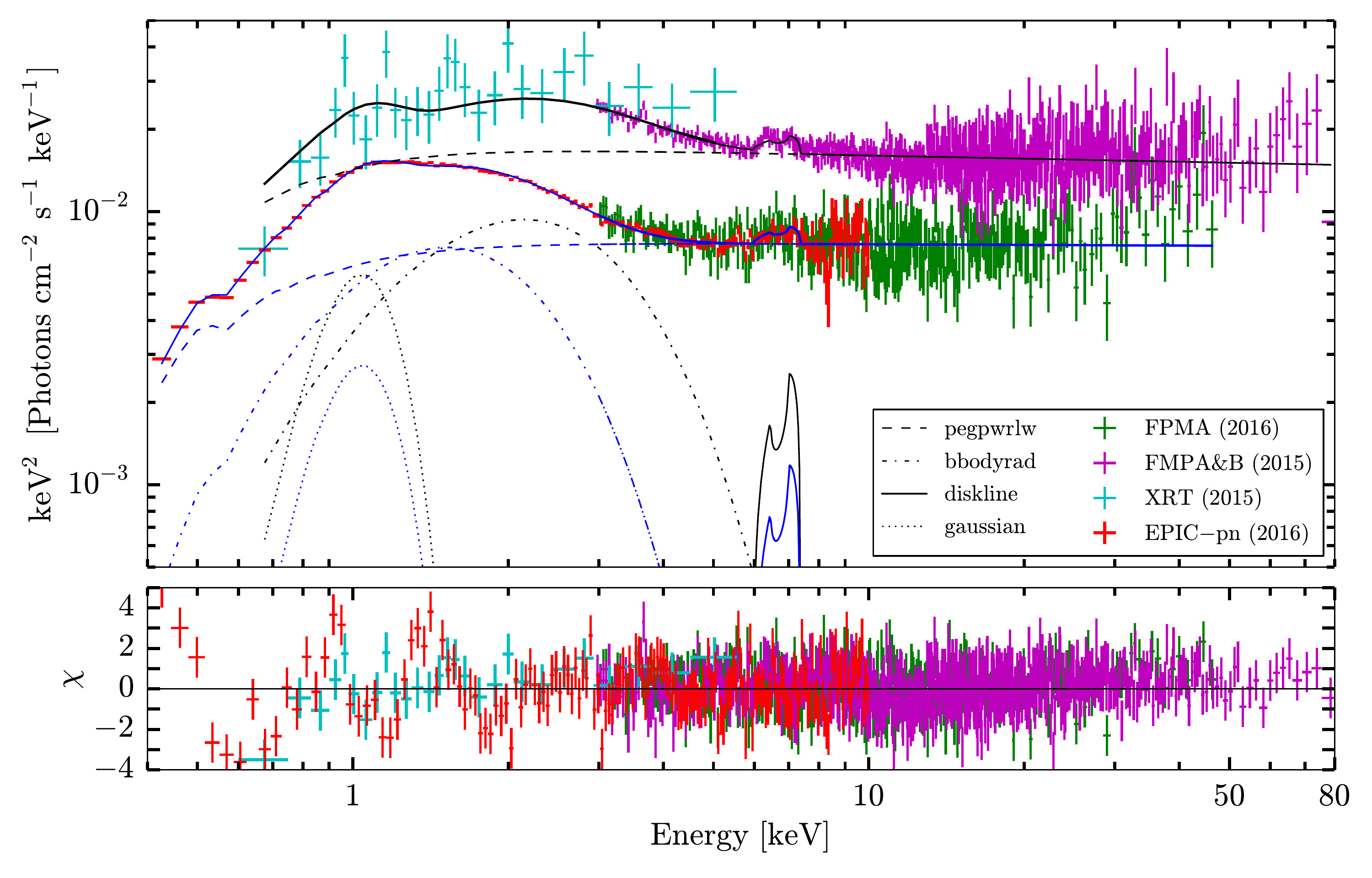}
    \caption{The best-fitting relativistic-reflecion model with the \textit{XMM-Newton} EPIC-pn, \textit{Swift}-XRT, and two \textit{NuSTAR} observations. \textit{Top:} all spectra unfolded around the best-fitting \textsc{constant*tbabs*(bbodyrad+pegpwrlw+gauss+diskline)} model. For visual clarity, the two sets of simultaneously observed spectra have seperately been rescaled by their fitted cross-calibration constant. For the same reason, we also do not plot the 2016 FPMB spectrum, which is consistent with the FPMA data. See Section \ref{sec:relref} for details on which parameters were tied between the two observational epochs. \textit{Bottom}: Residuals of the best fitting relativistic reflection model. Note that despite the Gaussian component, residual structure remains around $\sim 1$ keV, which we investigate in detail in Section \ref{sec:highres}.}
    \label{fig:spec2}
  \end{center}
\end{figure*}

All parameters are listed in Table \ref{tab:pars}, and the unfolded spectrum, best-fitting model and its residuals are plotted in Figure \ref{fig:spec2}. We note that the \textsc{diskline} model does not provide a better fit to the 2015 and 2016 data simulataneously than a simple Gaussian line. This is not entirely surpising as the large truncation radius implies a smaller distortion of the iron line shape by relativistic effects.

The bottom panel of Figure \ref{fig:spec2} shows that significant residuals remain below $2$ keV. As can also be seen in the bottom two panels of Fig. \ref{fig:spec1}, including a Gaussian feature around $1$ keV improves the model fit, but does evidently not describe the feature completely. To test the effect of this residual structure, we have refitted the full model excluding energies below $2$ keV. We only find significant changes in the parameters of the \textsc{bbodyrad} and the \textsc{gaussian} components. This is unsurprising, as the part of the spectrum described by these components is now removed. All other parameters remain unchanged. Interestingly, we also find that excluding the data below $2$ keV yields a $\chi^2_{\nu}$ of $1360.95/1368 \approx 1.00$ for the remaining data. Hence, we are confident that the residuals below $2$ keV do not influence our model fit. We will discuss these residuals in more detail in Section \ref{sec:highres}.

\begin{table*}
 \begin{center}
 \caption{\small{Best fit parameters to the \textit{NuSTAR} combined FPMA\&B (2015), \textit{Swift}-XRT (2015), FPMA and FPMB (2016), and \textit{XMM-Newton} EPIC-pn (2016) spectra. For parameters unlinked between the 2015 and 2016 datasets, the year is noted with the model component. The continuum model does not contain any Fe-K$\alpha$ component. Notes: *frozen. $^{\rm a}$flux between $0.3$ and $79$ keV. $^{\rm b}$emissivity index.}}
  \label{tab:pars}
   \begin{tabular}{llcccc}
  \hline \hline
  Component & Parameter [Unit] & Continuum & \multicolumn{3}{c}{Continuum + \textsc{diskline}} \\
  & & & $i=65^{\rm o}$ & $i=45^{\rm o}$ & $i=25^{\rm o}$ \\ \hline
  \textsc{tbabs} & $N_H$ [$10^{22}$ $\rm cm^{-2}$]      & $0.119 \pm 0.003$ & $0.118 \pm 0.003$ & $0.118 \pm 0.003$ & $0.116 \pm 0.003$ \\
  \textsc{pegpwrlw} (2015) & $\Gamma$                   & $2.08 \pm 0.01$ & $2.04 \pm 0.01$ & $2.04 \pm 0.01$ & $2.04 \pm 0.01$ \\
  & Norm [$10^{-12}$ erg cm$^{-2}$ s$^{-1}$]$^{\rm a}$        & $156 \pm 2$ & $147.5 \pm 2.1$ & $147.8 \pm 2.1$ & $147.2 \pm 1.9$ \\
  \textsc{pegpwrlw} (2016) & $\Gamma$                   & $2.00 \pm 0.01$ & $2.01 \pm 0.01$ & $2.01 \pm 0.01$ & $2.00 \pm 0.01$ \\
   & Norm [$10^{-12}$ erg cm$^{-2}$ s$^{-1}$]$^{\rm a}$        & $156 \pm 2$ & $147.5 \pm 2.1$ & $147.8 \pm 2.1$ & $147.2 \pm 1.9$ \\
  \textsc{bbodyrad} (2015) & $kT_{\rm BB}$ [keV]        & $0.47 \pm 0.03$ & $0.54 \pm 0.03$ & $0.54 \pm 0.03$ & $0.52 \pm 0.02$ \\
  & Norm                                                & $42.5^{+20.3}_{-13.6}$ & $23^{+8}_{-6}$ & $24^{+8}_{-6}$ & $28^{+10}_{-7}$ \\
  \textsc{bbodyrad} (2016) & $kT_{\rm BB}$ [keV]        & $0.364 \pm 0.003$ & $0.368 \pm 0.003$ & $0.368 \pm 0.003$ & $0.370 \pm 0.003$ \\
  & Norm                                                & $206.8 \pm 6.5$ & $189.2 \pm 6.3$ & $189.7 \pm 6.3$ & $186.9 \pm 6.2$ \\
  \textsc{gauss} & $E_{\rm centroid}$ [keV]             & $0.964 \pm 0.004$ & $0.961 \pm 0.004$ & $0.961 \pm 0.004$ & $0.959 \pm 0.005$\\
  & $\sigma$ [keV]                                      & $0.18 \pm 0.01$ & $0.18 \pm 0.01$ & $0.18 \pm 0.01$ & $0.19 \pm 0.01$\\
  & Norm [photons cm$^{-2}$ s$^{-1}$]                   & $(3.3 \pm 0.2)\times10^{-3}$ & $(3.3 \pm 0.2)\times10^{-3}$ & $(3.3 \pm 0.2)\times10^{-3}$ & $(3.4 \pm 0.2)\times10^{-3}$\\
  \hline 
  \textsc{diskline} & $E_{\rm line}$ [keV]              & -- & $6.73^{+0.06}_{-0.05}$ & $6.75 \pm 0.05$ & $6.97_{-0.02}$ \\
  & $q^{\rm b}$                                               & -- & $3$* & $3$* & $3$* \\
  & $R_{\rm in}$ [$R_g$]                                & -- & $77^{+22}_{-18}$ & $49^{+12}_{-11}$ & $6.45^{+0.68}_{-0.45}$ \\
  & $R_{\rm out}$ [$R_g$]                               & -- & $500$* & $500$* & $500$* \\
  & Norm [photons cm$^{-2}$ s$^{-1}$]                   & -- & $(4.7 \pm 0.5)\times10^{-5}$ & $(4.5 \pm 0.5)\times10^{-5}$ & $(5.9 \pm 0.6)\times10^{-5}$\\ 
  \hline 
  & $\chi^2/\rm d.o.f$ & $1705.4/ 1433$ & $1580.8/1430$ & $1582.7/1430$ & $1587.9/1430$ \\ \hline
  \end{tabular}
  \end{center}
\end{table*}

As stated, the inclination of the \textsc{diskline}-model is poorly constrained: all values between $5^{\rm o}$ and $90^{\rm o}$ lie within $3\sigma$ (e.g. $\Delta \chi^2 \leq 9$). Explicitly stepping through a grid in inclination and inner radius reveals a complicated $\chi^2$ space, where a high inclination and a truncated disk minimizes $\chi^2$ but a second, isolated minimum exists at an inclination of $\sim 25^{\rm o}$ and an inner radius around the ISCO. Hence, we cannot exclude a disk viewed at low inclination extending to the ISCO. We will discuss this further in Section \ref{sec:magnetictruncation}. 

For clarity, in Table \ref{tab:pars}, we also show the fit parameters for fixed inclinations of $45^{\rm o}$ (yielding $\Delta \chi^2 = +1.85$ and $R_{\rm in} = 49^{+12}_{-11}$ $R_g$) and $25^{\rm o}$ (yielding $\Delta \chi^2 = +7.05$, $R_{\rm in} = 6.45^{+0.68}_{-0.45}$ $R_g$). In the latter, the iron line energy peggs at its maximum value of $6.97$~keV. The continuum parameters do not change significantly with the inclination. Similarly, including the RGS spectrum does not influence either the parameters or the significances quoted in this and the previous paragraphs. 

Using the \textsc{relxill} reflection model, \citetalias{degenaar17} found a similarly truncated accretion disk in the 2015-only data, although the ISCO could not be excluded at $3\sigma$. When instead modelling the 2015 data with \textsc{diskline} reflection, we can exclude the ISCO at a significance of $\sim 4\sigma$. Thus, the addition of the 2016 observations allows us to more confidently infer that the inner disk in \source~is truncated away from the ISCO. 

\subsubsection{Broadband reflection: \textsc{relxill} and \textsc{reflionx}}
\label{relref}
A full relativistic reflection spectrum does not only consist of the Fe K$\alpha$ line, but contributes to the complete X-ray continuum, for instance through the presence of a Compton hump peaking around $10$--$20$ keV. Hence, we extended our analysis from the \textsc{diskline} reflection model to self-consistent models of the complete relativistically smeared reflection spectrum. We considered two options: (1) \textsc{relxill} \citep{dauser14,garcia14}, which models the illuminating powerlaw component simultanously with the reflection and thus replaces \textsc{pegpwrlw}, and (2) \textsc{reflionx} \citep{ross05} convolved with the \textsc{relconv}-model. In the second option, the illuminating flux is provided by the \textsc{pegpwrlw}-component in the continuum, whose power-law index is thus linked to the reflection spectrum. In both models, we again fixed the dimensionless spin $a$ to zero, the inclination to $65^{\rm o}$ and assume an unbroken emissivity profile with index $q=3$, consistent with both theoretical predictions \citep{wilkins12} and observations \citep[e.g.][]{cackett10}. Finally, we set the iron abundance to one and initially linked all reflection parameters between the 2015 and 2016 observations. Note that the untied continuum parameters (in the \textsc{blackbody} and \textsc{pegpwrlw}) remained untied. 

\begin{figure*}
  \begin{center}
    \includegraphics[width=\textwidth]{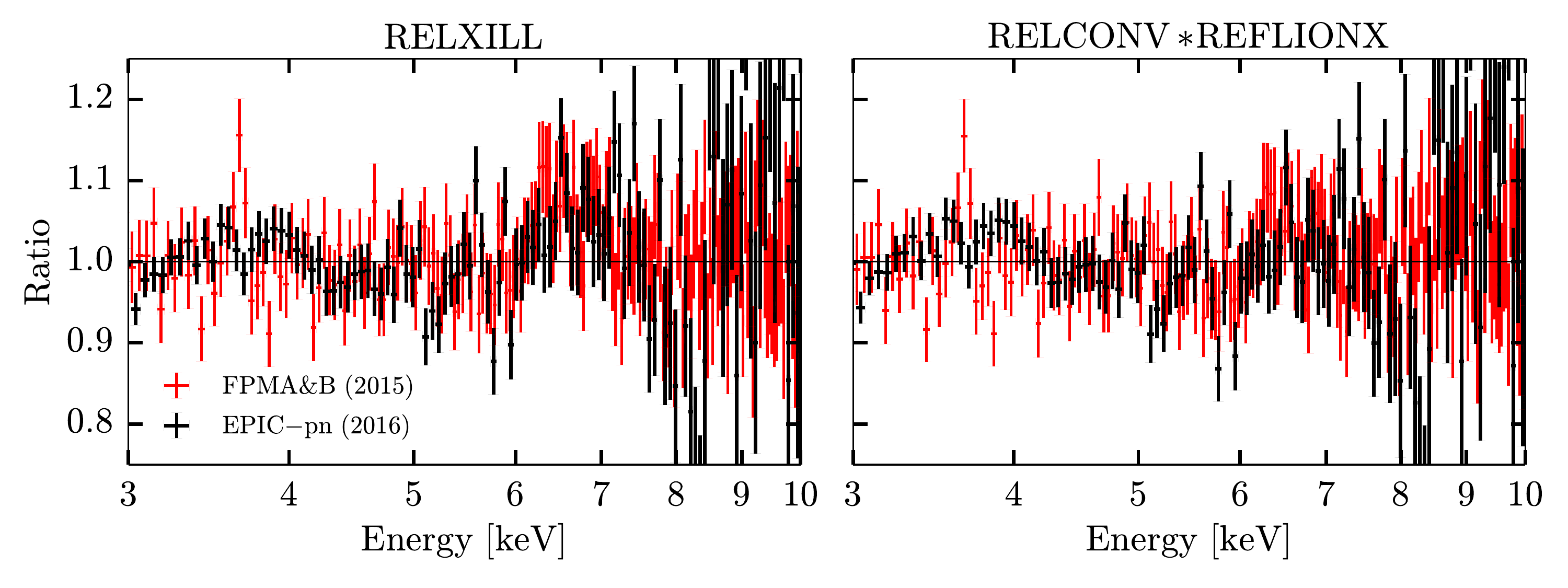}
    \caption{Data-to-model ratios for the \textsc{relxill} (left) and \textsc{reflionx}-based (right) broadband reflection models. The red data in both panels is the combined 2015 \textit{NuSTAR} spectrum, while the black data is the 2016 \textit{XMM-Newton} EPIC-pn spectrum. In both cases, clear iron-line residual structure remains between $\sim 6$ and $7$ keV.}
    \label{fig:fullrelref}
  \end{center}
\end{figure*}

Both broadband relativistic reflection models are unable to describe the 2015 and 2016 observations simultaneously with physically realistic parameters. The first model, using \textsc{relxill}, yields a $\chi^2_{\nu}$ of $1670.9/1430 = 1.17$ ( $\Delta \chi^2 \approx +90$ compared to the best \textsc{diskline}-models for the same number of free parameters). Importantly, the reflection parameters are ill-constrained: the inner radius pegs at the minimum value of $6$ $R_g$, but all values up to $120$ $R_g$ are consistent within $3\sigma$. Additionally, the iron line complex is badly modelled: Figure \ref{fig:fullrelref} (left) shows the residuals between $3$--$10$ keV, showing clear residual iron line structure around $6.5$ keV. Similar problems arise for the second, \textsc{reflionx}-based model. While the quality of the fit is slightly better ($\chi^2_{\nu} = 1602.3/1430 = 1.12$), the inner radius is again unconstrained: its best fit value is $\sim 395$ $R_g$, while the outer disk radius was fixed to $400$ $R_g$, and all values down to the ISCO are consistent within $3\sigma$. Furthermore, again clear residual structure remains in the data-to-model ratio, as can be seen in the right panel of Figure \ref{fig:fullrelref}.  

Despite the similarities between the 2015 and 2016 spectra, fitting both simultaneously with a broadband reflection spectrum could be the cause of the problems detailed above. However, untying the reflection parameters between the two sets of observations does not resolve those issues. In the \textsc{relxill}-model, this results in a marginally significant improvement ($\Delta \chi^2 \approx -37$ for $4$ additional degrees of freedom, f-test rejection probability $\sim 2\times10^{-6}$), but the inner radii remain unconstrained and the residual structure does not disappear. For the \textsc{reflionx}-models, untying the reflection parameters does not result in a significant improvement ($\Delta \chi^2 \approx 13$ for $\Delta \rm d.o.f. = 3$, f-test rejection probability $p \sim 0.009$), while the two inner radii both exceed $400$ $R_g$. Finally, the same iron-line structure in the data-to-model ratio remains. For both models, we also attempted a broken emissivity profile with $q_1=0$ and $q_2=3$, which is more appropriate for a large scale height of the corona -- again, this offered no improvements to the modelling. 

Based on the above considerations, we conclude that the data quality of our 2016 observations is not sufficient to constrain the full broadband relativistic reflection spectrum. \citetalias{degenaar17} were able to model the 2015 observation using \textsc{relxill}, although the inferred inner radius was barely constrained; while $R_{\rm in}$ was found to exceed $100$ $R_g$, the ISCO could not be excluded. Even though we fixed several parameters in our reflection fits (among others spin and inclination), the lower flux in the 2016 observation does not allow us to apply a model more complicated than \textsc{diskline}. It should be noted again that in 2015, \source~was the first NS LMXB where the iron line could even be detected at these low fluxes. Hence, it is not surprising that the data does not allow for the most detailed analysis of the reflection.

\subsubsection{The $\sim 1$~keV excess}

Finally, we briefly discuss the $\sim 1$~keV excess emission. Similar soft excesses have been observed in the fast modes of the EPIC-pn instrument (Guainazzi et al. 2014, XMM-SOC-CAL-TN-0083\footnote{\href{http://xmm2.esac.esa.int/external/xmm_sw_cal/calib/documentation/index.shtml}{http://xmm2.esac.esa.int/external/xmm\_sw\_cal/calib/doc} {\color{blue}umentation/index.shtml}}). However, this instrumental effect is typically observed in highly obscured sources. As $N_H$ is a factor $\gtrsim 5$ lower for \source~than the sources where this issue is reported, we do not expect that this effect plays a role (see for instance \citet{hiemstra11}, where the $N_H$ is $\sim 65$ times higher). As discussed later, we also observe a similar feature in the RGS spectrum, strengtening the case that the feature is real. 

Alternatively, it could arise from reflection. In addition to the iron line and the Compton hump, reflection can provide a significant contribution around $1$ keV. As suggested by \citetalias{degenaar17} and the failure of broadband reflection models to describe the data, the $\sim 1$ keV excess might thus originate from a second, more distant reflection site. Hence, we adjusted the best fitting \textsc{diskline}-model, replacing the $\sim 1$~keV Gaussian component with a second, unlinked \textsc{diskline} component. However, this results in a significantly worse fit, with $\Delta \chi^2  \approx +320$ for the same number of free parameters. Moreover, the second reflection site would be located at $\sim 17$ $R_g$, which is within the truncation of the accretion disk inferred from the iron line. Hence, we do not find evidence for a second reflection site from the EPIC-pn data. 

\section{High-resolution spectroscopy}
\label{sec:highres}

In the 2014 \textit{Chandra}-HETG observation of \source, \citetalias{degenaar17} detected several marginally significant emission and absorption lines, possibly originating from a outflow. However, the unambigious identification of the lines and their origin proved difficult based on the \textit{Chandra} data alone. Our \xmm EPIC-pn spectrum contains a clear excess around $0.9$--$1.0$ keV ($\sim 12$--$13$\AA). In order to investigate the nature of this excess and revisit the detection of the narrow lines in the \textit{Chandra} spectrum, we perform a high-resolution spectral analysis of the \xmm RGS spectrum. In this section, we first discuss the RGS continuum, followed by an initial phenomenological line search and subsequent physical modelling. In this section, we switch from energy in keV to wavelength in \r{A}ngstr\"om, as is common in high-resolution X-ray spectroscopy.

\subsection{RGS continuum}
\label{sec:rgscont}

Before focussing on narrow lines and the $\sim 1$ keV excess, we investigated the properties of the RGS continuum. The $\sim 1$~keV ($\sim 12.4$ \AA) excess emission in the EPIC-pn spectrum is described with a simple Gaussian in the previous section, but the bottom panel of Figure \ref{fig:spec2} shows that this is not fully adequate. Figure \ref{fig:lineresults} (top panel) shows the first and second order RGS spectra, unfolded around a constant. An emission excess is visible around $11$--$12$ \AA, together with a strong oxygen edge around $23$ \AA. The neon edge around $14.2$ \AA, though, appears not as strong as the oxygen edge. In a number of bins, the first and second order spectra deviate more than the uncertainties and hence the results of more detailed line-modelling (Section \ref{sec:linemodelling}) should be interpreted with caution. 

We first attempted to describe the RGS continuum with a simple absorbed blackbody model. However, such a model does not provide a good description of the data; the first order spectrum is ill-described around and above the oxygen edge. Although the blackbody temperature of $T_{\rm BB} \approx 0.35$ keV is consistent with the broadband spectral analysis, the hydrogen column density $N_H \approx 0.35\times10^{21}$ cm$^{-2}$ is a factor three below typically observed values and predictions based on ISM maps \citep{kalberla05}. Including a \textsc{powerlaw}-component, with $\Gamma$ fixed to a value of $2.05$ since it is ill-constrained at these low energies, significantly improves the continuum description: $\chi^2_{\nu} = 946.30/768 = 1.23$ (f-test rejection probability $p \sim 10^{-13}$). While the resulting $N_H \approx 1.2\times10^{21}$ cm$^{-2}$ and $T_{\rm BB} \approx 0.33$ keV are in line with the full spectrum, the discrepancies around and above the oxygen edge largely remain (see the blue model in Figure \ref{fig:lineresults}, top panel). 

To more accurately model the oxygen edge in the continuum, we replaced the simple absorption model \textsc{tbabs} by the more detailed \textsc{tbnew}-model. The \textsc{tbnew}-model allows the absorption abundances of individual species to vary with respect to the Solar abundances by \citet{wilms00}. We fixed the value of $N_H$ to $1.2\times10^{21}$ cm$^{-2}$ \citep{kalberla05} and first allowed oxygen to vary. This model results in a significantly improved fit of the order 1 and 2 RGS spectra ($\chi^2_{\nu} = 856.36/767 = 1.12$, f-test rejection probability $p \sim 10^{-18}$) for a high oxygen abundance of $A_{\rm O} = 1.94$, and a blackbody temperature of $T_{\rm BB} \approx 0.31$ keV. Figure \ref{fig:lineresults} shows this continuum model both with solar abundances and an enhanced oxygen abundance; the discrepancies between the two models above $18$ \AA~are evident. Alternatively, instead of being due to a an enhanced oxygen abundance, the excess emission above the oxygen edge might result from a combination of many C and N lines. However, such lines are clearly not resolved and an enhanced oxygen abundance is sufficient to model to full continuum. 

In order to understand the nature of the high oxygen abundance, we also attempted to free the magnesium, iron and neon abundances, for both a fixed and a free oxygen abundance. None of these options resulted in a significant improvement of the fit. It appears that, with the exception of oxygen, all absorption edges are correctly modelled by the (fixed) interstellar value of the hydrogen absorption. This implies that the high oxygen abundance originates from the source -- if it were interstellar, a similar increase would be expected for other abundances, such as neon. Secondly, the oxygen abundance in the ISM is not expected to deviate from the solar value by more than a factor $\sim 1.3$ \citep{pinto13}. Hence, the additional oxygen absorption might instead have a circumbinary origin, as we will discuss in Section \ref{sec:ucxb}.

\subsection{Line search}

After analysing the continuum properties in the RGS spectra, we turned to an explorative search for narrow emission and absorption lines. Following the method detailed in \citet{pinto16}, we adopted the continuum model and subsequently added a narrow gaussian line with a fixed width of $500$ or $2000$ km s$^{-1}$. We then fitted the normalisation of this gaussian line, calculated its error, shifted the line by $0.01$ \AA~and repeated. This procedure returns, at each gridpoint in wavelength, two indications for the presence of a narrow line: the line normalisation divided by its error, and the improvement in the C-statistic $\Delta C$. Note that we employ the C-statistic instead of $\chi^2$-statistics for the detailed line search and the subsequent line modelling, as it is more accurate for low counts per bin. We stress that both measures are \textit{single trial} estimations of the significance of the narrow line; both merely hint to the presence of emission or absorption but can be prone to false positives when considering only a single dataset. Hence, the comparison with the similar \textit{Chandra} line search in \citetalias{degenaar17} is essential to rule out possible false positives.

The order 1 and order 2 RGS spectra were fitted simultaneously and searched in the ranges $7$--$28$ \AA~and $7$--$16$ \AA, respectively, where the source is significantly detected above the background. We excluded the range below $7$ \AA, as calibration issues between the first and second order detectors result in large discrepancies between the two spectra. We explicitly checked whether freezing the continuum parameters influences the line search, but found that this does not alter the outcome. 

\begin{figure*}
  \begin{center}
    \includegraphics[width=\textwidth]{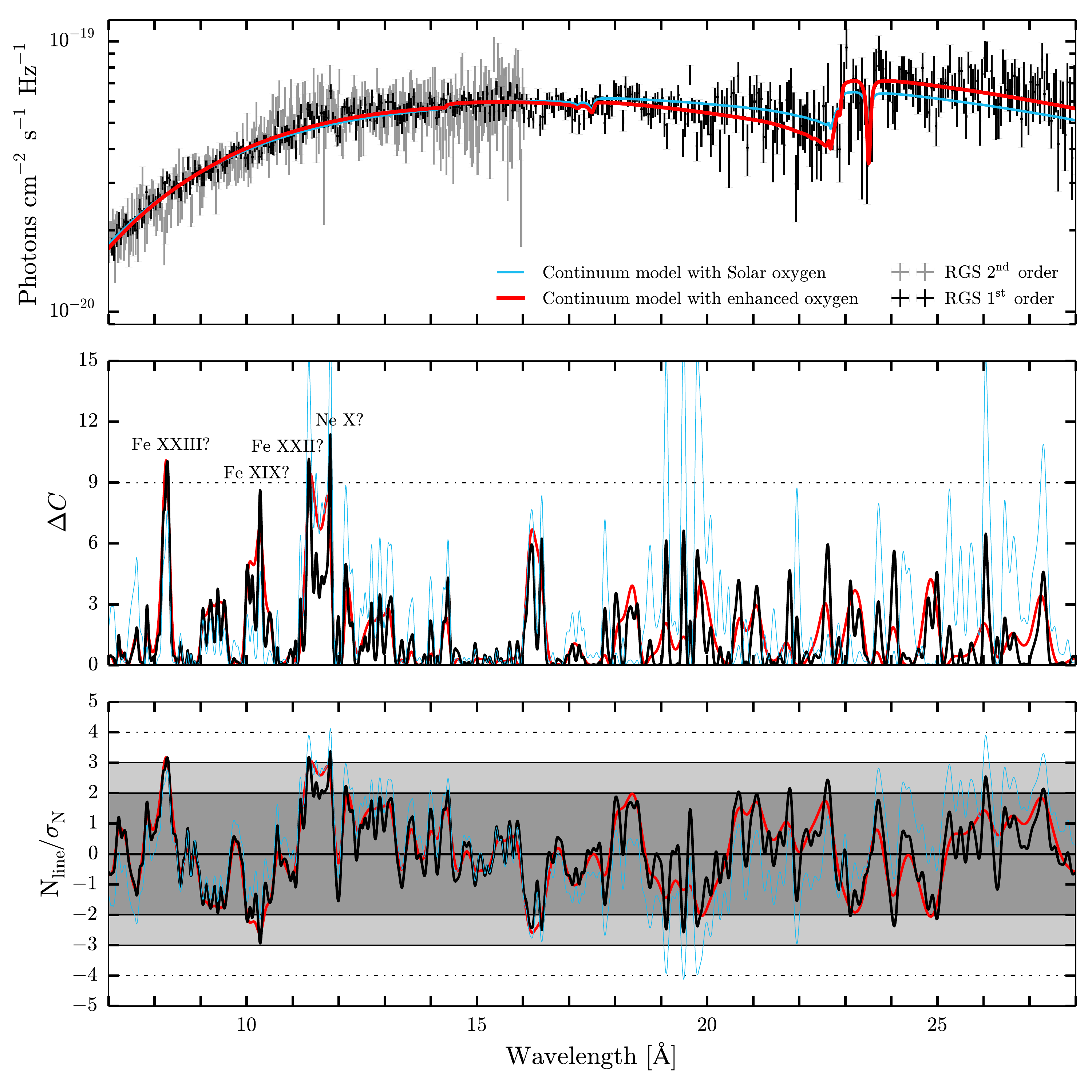}
    \caption{Results of the narrow line search in the RGS spectra for two different continuum models. The top panel shows the RGS spectra and the two continuum models, all unfolded around a constant to remove the instrument response. The middle panel shows the improvement in C-statistic, for the addition of one free parameter (the line normalisation). The dashed line indicates $\Delta C=9$, corresponding to a $3\sigma$ single-trial significance. The bottom panel shows the fitted line normalisation divided by its $1\sigma$-uncertainty, where a positive normalisation implies emission and a negative one absorption. Different colors correspond to different continua: the black and red curves are calculated with the \textsc{tbnew*(bbodyrad+po)}-model, assuming free O and respectively a $500$ and $2000$ km s$^{-1}$ linewidth. The blue curve corresponds to a model with Solar abundances in the absorption column; the remaining trends in $N_{\rm line}/\sigma_N$ reveal the need for a non-Solar oxygen abundances.}
    \label{fig:lineresults}
  \end{center}
\end{figure*}

Figure \ref{fig:lineresults} shows the results of our phenomenological search for narrow lines: the middle panel shows the decrease in the C-statistic $\Delta C$, where $\Delta C = 9$ corresponds to a 3-sigma single-trial improvement. The bottom panel shows the line normalisation divided by its error, where again $N_{\rm line}/\sigma_{\rm N} = \pm 3$ indicates the 3-sigma single-trial significance level. The black and red thick curves show the results for a linewidth of $500$ km s$^{-1}$ and $2000$ km s$^{-1}$, assuming a continuum with enhanced oxygen: both velocities return consistent results, showing several possible emission features and a single possible absorption feature. Note that the 2 potential emission lines around $11$--$12$ \AA~are within the puzzling $\sim 1$ keV excess observed in the EPIC-pn spectrum. Finally, we also show the results assuming solar abundances in blue: clear residual trends in the bottom panel remain, as $N_{\rm line}/\sigma_{\rm N}$ generally slopes downwards between $12$--$20$ \AA~and upwards between $20$--$28$ \AA, artificially enhancing the significances of any lines.    

The line search returns three emission lines, at $8.3$, $11.35$ and $11.8$ \AA, with at least a $3\sigma$ single trial significance. Interestingly, similar lines are observed by \citetalias{degenaar17} in the \textit{Chandra} spectrum, strenghtening the case that these are physical. Comparing both line searches, the emission lines are possibly associated with blueshifted Fe XXIII (restframe $8.82$ \AA), Fe XXII-XXIII ($11.75$ \AA) and Ne X ($12.125$ \AA), respectively. The corresponding blueshifts, ranging from $z \sim -0.03$ to $z \sim -0.06$, are comparable although not fully consisent. We also see an absorption line at $10.29$ \AA, as was also found by \citetalias{degenaar17}, which is consistent with Fe XIX (restframe $10.82$ \AA) blueshifted by $z \sim -0.05$. Additionally, hints of a broad emission feature between $18$--$18.5$ \AA~can be seen in the top panel of Figure \ref{fig:lineresults}. While it is not picked up as a narrow line in the search, the position is consistent with a combination of blueshifted O VII and OVIII lines. If so, the blueshift of the OVIII line would lie in the range $z\sim -0.03$ to $\sim -0.05$. 

We do not confirm several (hints of) absorption lines seen in \textit{Chandra}. This could arise due to differences between the detectors (for instance the low efficiency of RGS compared to HETG around $7.5$ \AA) or differences is the used continuums: \citetalias{degenaar17} did not use an enhanced oxygen abundance, which can result in the artificial enhancement of the line search significances. Finally, some of the \textit{Chandra} lines could of course also simply be statistical fluctuations. 

\subsection{Line modelling}
\label{sec:linemodelling}

The phenomenological line search hints towards the presence of a handful of narrow absorption and emission lines in the RGS spectra. In order to further investigate the nature of these lines and the $\sim 1$ keV excess, we applied two different types of line models on top of the continuum model: (1) \textsc{bapec}, a collisional ionisation model expected for a shock origin, such as in a jet, and (2) \textsc{photemis}, a photo-ionisation model more suggestive of a wind origin. We assumed no velocity line broadening. Since the abundances remained unconstrained when left to vary, we also assumed Solar abudances in both models, despite the enhanced oxygen abundance in the absorption column. Fixing these two parameters helps by reducing the number of free and possibly degenerate parameters. In both line models, we initially set the redshift parameter to zero, and subsequently let it vary between $-0.2$ and $0.2$. We also let the continuum parameters, except for $N_H$, free to vary. We employ C-statistics and the initial continuum C-value is $C_{\rm cont} = 862.84$ for $767$ degrees of freedom. 

First, we applied the collisional ionisation model \textsc{bapec} on top of the continuum. Assuming no redshift, we find an improvement of the fit of $\Delta C \sim 29$ for two additional parameters: the normalisation and the temperature. Subsequently varying the redshift between $-0.2$ and $0.2$ results in the best line-model fit, with $C_{\rm bapec} = 807.76$ for $764$ parameters ($\Delta C = 55.08$ with respect to the continuum). We find a temperature of $kT_{\rm bapec} = 1.15^{+0.06}_{-0.07}$ keV and a blueshift of $z = -0.048 \pm 0.001$, corresponding to $\sim 15000$ km s$^{-1}$. We do note that a number of additional local minima are located at different combination of redshift and temperature. However, these show a significantly lower change in C-statistic of maximally $\Delta C \sim 40$. 

\begin{figure}
  \begin{center}
    \includegraphics[width=\columnwidth]{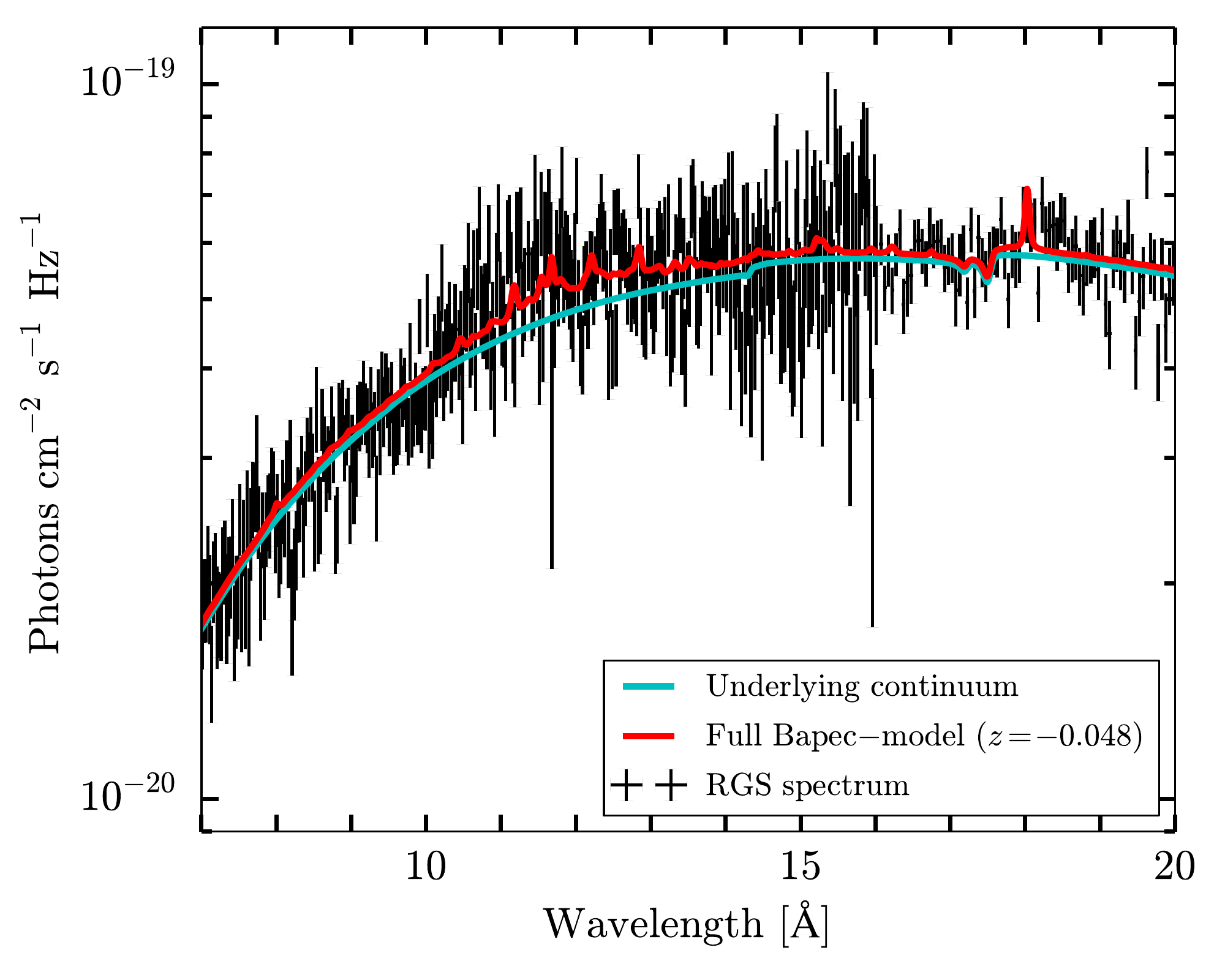}
    \caption{The \textsc{bapec}-linemodel fit. The red line shows the complete best-fit model with $z = -0.048$, while they cyan line shows only the continuum of the best model fit. The line model fits both narrow lines and part of the continuum, most dominantly in the region around the EPIC-pn excess at $\sim 12$--$13$\AA.}
    \label{fig:linemodel}
  \end{center}
\end{figure}

In Figure \ref{fig:linemodel}, we show the RGS spectra, the underlying continuum model and the best-fitting \textsc{bapec} model. The \textsc{bapec}-model fits both narrow lines and the $\sim 1$ keV excess, the latter with a pseudo-continuum of weak lines. In addition, it also accounts for the emission excess around $18$~\AA. However, there also appear to be small discrepansies between the position of the narrow lines in the model and the data, that we will discuss in Section \ref{sec:caveats}. Finally, we attempted the addition of a second \textsc{bapec}-component, with the same temperature and normalisation but opposite velocity, mimicking the emission from second, receding outflow. This results in a comparable fit with $C=802.80$ and a slightly lower red/blueshift of $z\sim \pm0.035$ for the two \textsc{bapec}-components.

\begin{figure}
  \begin{center}
    \includegraphics[width=\columnwidth]{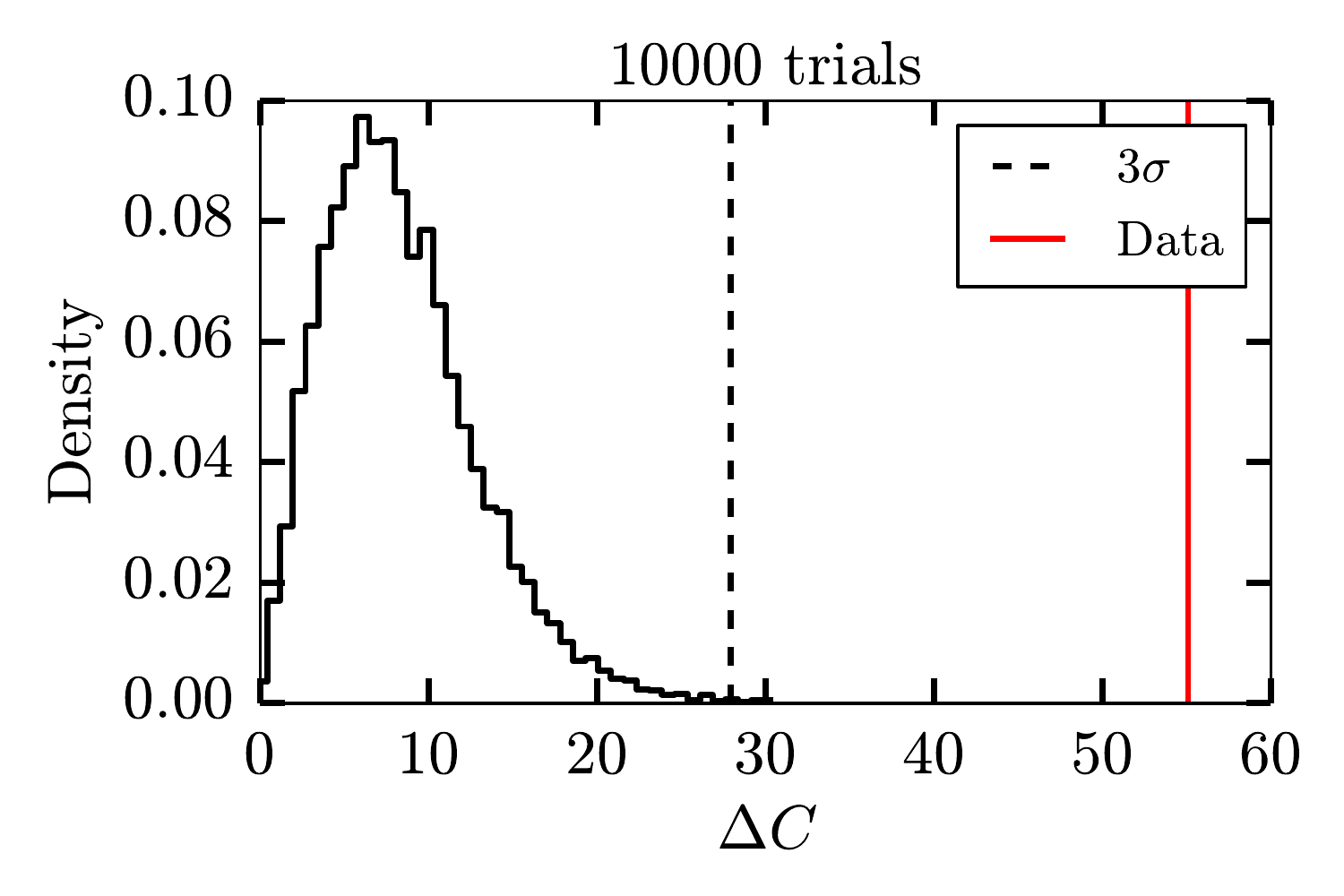}
    \caption{Histogram of results of the Monte-Carlo simulation of the \textsc{bapec}-linemodel significance. $\Delta C$ is the chance improvement of the C-statistic when fitting the linemodel to simulated spectra without any narrow lines. The red line indicates the measured $\Delta C$ in the observed RGS spectra of \source}
    \label{fig:dC}
  \end{center}
\end{figure}

We performed a Monte-Carlo simulation to check the significance of the \textsc{bapec}-component and test whether its presence could result from a statistical fluctuation. For this purpose, we simulated $10^4$ sets of first and second order RGS spectra from the best fit \textit{continuum} model, an exposure time of $\sim 127$ ks and the observed backgrounds. The exposure time accounts for the combination of the seperate spectra from each detector, with individual exposures of $\sim 63.5$ ks, into one spectrum per order. We then fit the fake spectra, simulated \textit{from the continuum only}, first with the continuum model and afterwards with the continuum plus \textsc{bapec} model. Finally, we save the change in fit statistic between the two fits. In Figure \ref{fig:dC}, we plot a histogram of the resulting $\Delta C$ values. The value of $\Delta C$ in our real observations evidently greatly exceeds any of the values from the simulated spectra. While the calculated number of trials ($10^4$) formally yields a $3.7\sigma$ significance of the \textsc{bapec}-component, we note that \textit{none} of the trials exceeded, or even approached, the observed $\Delta C$ value. It is important to reiterate that this significance is not merely due to the modelling of narrow lines but also largely due to the pseudo-continuum of weak lines fitting the broad $\sim 1$ keV excess.

Alternatively, the photo-ionisation model \textsc{photemis} is not able to adequately model the lines and the broad excess in the RGS spectra; assuming zero red- or blueshift, the best fit results in an improvement of $\Delta C = 0.42$ for 2 extra free parameters (the normalisation and ionisation parameter). Freeing the redshift does not immediately improve the fit. As the \textsc{photemis} model appears to be relatively inefficient in finding the global minimum fit statistic, we also explicitly searched a grid in redshift and ionisation. Sampling the blueshift between $z = -0.2$ and $z = 0.0$ and ionisation parameters between $r\log\xi = 1.0$ and $r\log\xi = 4.0$, we find the best fit at $z \sim -0.164$ and $r\log\xi \sim 1.5$ with $\Delta C = 16.52$. However, this model does not adequately model the clear excess emission around $11$--$12$ \AA. Hence, we conclude that \textsc{photemis} can not adequately model the RGS spectra and that we do not observe hints for a photo-ionised wind in \source, as suggested by \citetalias{degenaar17}. This is consistent with the apparent stronger emission from Fe and Ne X compared to O VIII, as is expected in a plasma that is collisionally ionized instead of photo-ionized.

\citetalias{degenaar17} were able to describe the absorption features in the HETG spectra of J1706 using the \textsc{pion}-model in \textsc{spex}, which is the equivalent of \textsc{photemis}. This photo-ionized plasma model fit is primarily driven by a broad absorption feature around $15$-$16$\AA, which is not observed in our RGS spectra. As stated before, this difference might arise due to the difference in continuum modelling (i.e. including an enhanced oxygen abundance). Alternatively, the feature might be too broad and shallow to be picked up in our narrow line search and to be significant in the line model fits.

\subsection{Reflection modelling}

Although the EPIC-pn excess at $\sim 1$ keV can not be modelled with a \textsc{diskline}-model, we considered a reflection origin of the observed emission and absorption features in the RGS spectrum. We initially tried three different models: (1) \textsc{xillver}, which does not include relativistic blurring \citep{garcia13}, (2) \textsc{relxill}, which does include blurring, and (3) \textsc{diskline}. In the first two cases, the reflection model contains a powerlaw component. Hence, we both use this included powerlaw to model the continuum-powerlaw and add it on top of the complete continuum \citep[as in][]{madej14}. 

All five resulting combinations of continuum and reflection fail to model the observed narrow features in the RGS spectrum, as they tend towards high ionisations where neither narrow lines not broadended features are prominent; as a result, neither the emission features around $11$--$12$ \AA~nor those around $18$ \AA~are accounted for, the parameters remain unconstrained, and the reflection model simply mimicks the continuum powerlaw. This is not particularly surprising -- even the broadband spectra, that are more suitable for fitting the complete broadband reflection models, are too faint too adequately constrain such models despite the clear iron K$\alpha$ line.  

As a final check, we applied the \textsc{xillver}$_{\rm CO}$-model \citep{madej14}, which models reflection off an oxygen-rich disk in an UCXB. Given the recent evidence for a UCXB nature in \source~({\color{blue} Hern\'andez Santisteban et al., 2017.}) and the enhanced oxygen absorption in our continuum modelling, such a model might be more applicable. However, the same problems as above arise; we try adding the \textsc{xillver}$_{\rm CO}$-component to the full continuum model, and replacing the powerlaw-component by the reflection model, both with and without relativistic blurring. None of these options can either significantly improve the fit or account for any of the emission features between $11$--$12$ \AA. This is again not surprising, as the soft reflection features from a CO disk are expected between $15$--$20$ \AA~\citep[see][Figure 4]{madej14}. Hence, we find no evidence that these features arise from either the same reflection site as the iron K$\alpha$ line of a more distant site, as suggested by \citetalias{degenaar17}. 

\section{Timing analysis}
\label{sec:timing}

\citet{Strohmayer17} reported the detection of pulsations at $163.655$ Hz in the only \textit{RXTE} observation of \source, taken in 2008, making it the $19^{\rm th}$ discovered accreting milli-second X-ray pulsar \citep[AMXP; see e.g.][]{patruno12, patruno17}. The signal is detected in the $2$--$12$ keV energy band at a $4.3\sigma$ overall significance. Given the short exposure of the observation ($\sim 1$ ks), the orbit can only be constrained to $\gtrsim 17$ minutes, although a dynamical power spectrum does suggest an orbitally induced variation of $\Delta \nu \approx 0.002$ Hz. As our \textit{XMM-Newton} EPIC-pn observation is $\sim 63$ ks in timing mode, detecting the pulsation could provide us with an orbital solution and a confirmation of the AMXP nature of \source. For this purpose, we applied a simple FFT pulsation search, an acceleration search and a semi-coherent search of the \textit{XMM-Newton} observation. We explicitly checked the first two methods on the \textit{RXTE} observation as well, confirming the results by \citet{Strohmayer17}. 

We barycentered the photon arrival times using the \textsc{barycorr}-tool in \textsc{SAS} with the source position from \citet{ricci08}, and extracted light curves in the full $0.5$--$10$ keV and $2.0$--$10.0$ keV energy bands. Similar to what \citet{Strohmayer17} used for the RXTE data, we rebinned our XMM data to a time resolution of $2^{-13}$ s, corresponding to a Nyquist frequency of $4096$ Hz. We then FFT'ed the light curves and computed individual, Leahy-normalised power spectra of segments of length $64$, $128$, $256$, $512$ and $1024$ s (i.e. corresponding to a $1/64$ to $1/1024$ Hz frequency resolution). Given the frequency drift reported by \citet{Strohmayer17}, combined with the possible UCXB-nature of \source~({\color{blue} Hern\'andez Santisteban et al., 2017.}), we do not search longer segments: the orbital frequency drift would become large and spread out the signal over multiple frequency bins. For instance, in a $2048$ s segment, a signal with the reported drift of $\sim 0.002$ Hz ks$^{-1}$ would be divided over 8 bins. 

\begin{figure}
  \begin{center}
    \includegraphics[width=\columnwidth]{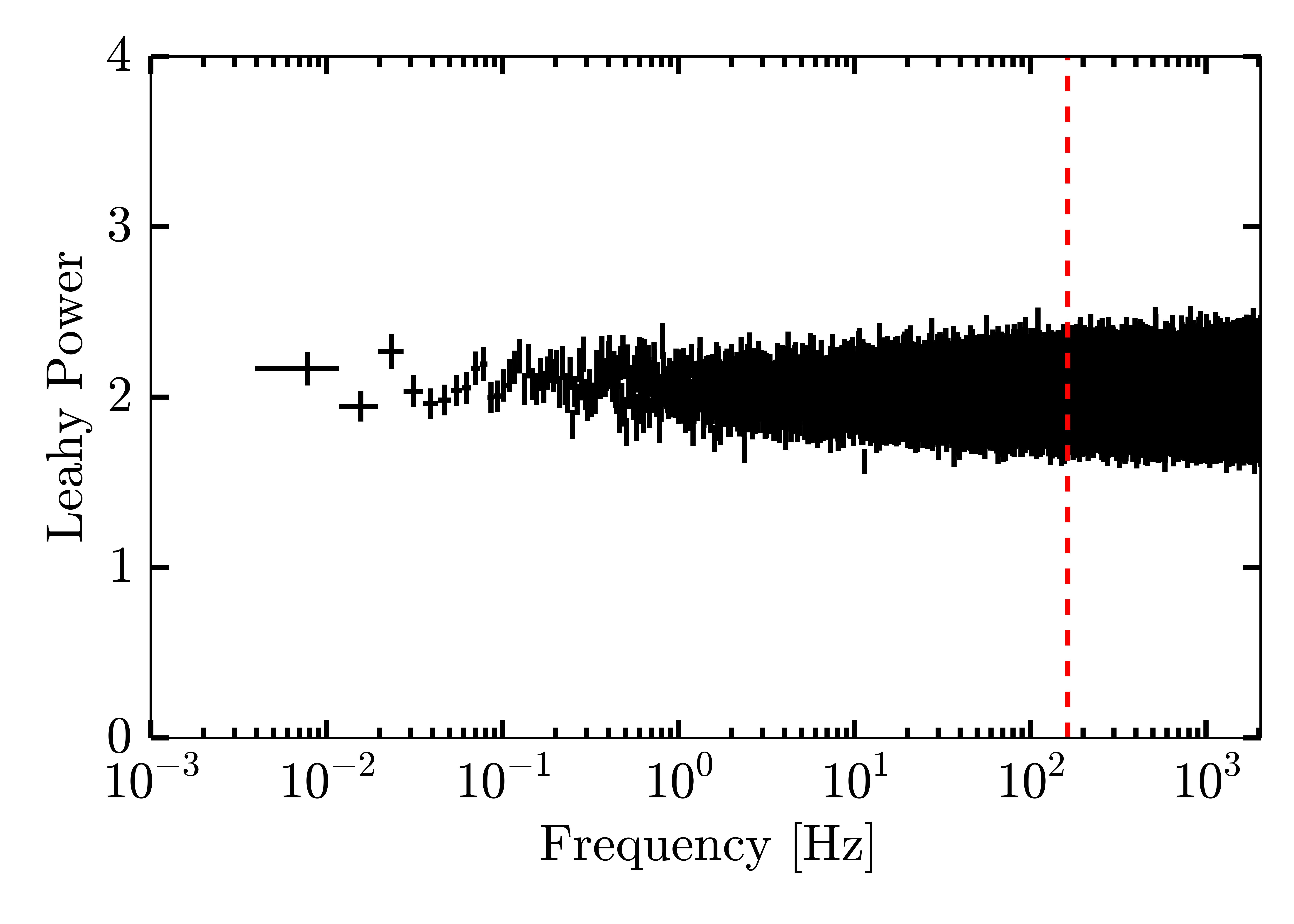}
    \caption{Leahy-normalised power spectrum of the \textit{XMM-Newton} observation of J1706. This power spectrum shows the average of all power spectra generated from $128$ s segments with a Nyquist frequency of $4096$ Hz. No pulsations are visible at the reported pulsation frequency, shown by the red dashed line.}
    \label{fig:psd}
  \end{center}
\end{figure}

We do not detect any significant pulsation at any frequency, including in the $163$--$164$ Hz range, in any individual power spectrum for both energy bands. The same holds when we average all power spectra computed from segments of the same length, in order to reduce the noise. We show an example of such an averaged power spectrum, using $128$ s segments, in Figure \ref{fig:psd}. The red line shows the pulsation frequency reported by \citet{Strohmayer17}. To overcome the trade-off between total counts (pushing a long segment size) and orbital frequency drift (pushing a short segment size), we apply two more sophisticated techniques with a higher sensitivity. 

Firstly, we applied an acceleration search using the \textsc{accelsearch} routine in \textsc{PRESTO}\footnote{\href{http://www.cv.nrao.edu/~sransom/presto/}{http://www.cv.nrao.edu/$\sim$sransom/presto/}}, described in detail in \citet{ransom02}. Here, the assumption is that over a small fraction of the orbit (maximally $\sim 10\%$), the orbital acceleration and thus the frequency drift is approximately constant. The smeared out pulse signal is recovered by combining the power in adjacent bins. As \textsc{PRESTO} was originally developed for radio data, we first converted the \textit{XMM-Newton} event tables to a binary file with the photon times of arrival using custom software. We computed such binary files for the same two energy bands as before; for each band, we analyse the entire observation (where the acceleration is definitely not constant) and individual $1000$ and $2000$~s segments. We focused the acceleration search on the range between $100$ and $200$ Hz, combining maximally 200 adjacent frequency bins to recover a signal. 

Again, no significant signal is present at $163$--$164$ Hz in any of the segments (full observation, $1000$ s, or $2000$ s) in either energy band. This lack of a significant signal is not necessarily surprising; given that the assumption of a constant orbital acceleration holds for approximately $10\%$ of the orbital period, an acceleration search in a $1000$ s segment is only effective for orbital periods of $\gtrsim 2.78$ hours. Instead, the orbit in \source~is likely to be shorter ({\color{blue} Hern\'andez Santisteban et al., in 2017.}). However, using shorter segments would reduce the signal to noise such that a signal might not be detected either. We tested this explicitly be checking $200$ s segments as well, where we do not find any (real or instrumental) features at a single-trial significance of $\geq 3\sigma$.

We do find signals at $130.57$ Hz and $125.13$ Hz recurring in several of the $1000$ and $2000$ s segments. To test their nature, we exactly repeated our analysis on \textit{XMM-Newton} EPIC-pn timing-mode observations of the NS LMXBs MXB 1730-335 (i.e. the Rapid Burster; obsid 0770580601) and HETE J1900.1-2455 (obsid 0671880101). In both sources, these signals are present as well, confirming that their origin is not related to J1706. 

Finally, we applied a semi-coherent pulsation search as described in detail in \citet{messenger11} and \citet{messenger15}, using the EPIC-on instrument time resolution of $29.56$ $\mu$s. In this method, a physically motivated section of the relevant parameter space (e.g. spin frequency, orbital period, semi-major axis and orbital phase) is sampled to produce a set of template orbits. The pulsations are then searched in two stages, a coherent one and an incoherent one. The coherent stage uses short data segments, in our case 41-s long, and searches for a coherent signal (i.e., tracking the signal phase) using a Taylor series expansion in frequency derivatives as the signal model. In the incoherent stage the coherent segments are combined and the signal power summed according to the template orbital and spin parameters.

In our search we explored spin frequency in the range 163.63--163.67$\rm\,Hz$ , orbital periods between 0.25 and 6 hours, a semi-major axis between 0.01--1 lt-s and an orbital phase $0$--$2\pi$. While centered around the expected values, the extent of the parameter space selected is not dictated by a true physical motivation but rather by the limited computational power used. This approach overcomes the limitation of the acceleration search, where the low count rate makes finding signal in short segments challenging: in the semi-coherent search, the entire observation is analysed by explicitly including the non-linear orbital frequency drift in the analysis. We find, however, no signal in the \textit{XMM-Newton} data, with 90\% false alarm probability upper limits on the pulsed fraction of $5.4\%$. An additional search covering an expanded spin frequency range of 158.655--168.655 Hz, at ~30\% lower sensitivity, also failed to detect any significant signal.

\section{Discussion}
\label{sec:discussion}

We present an extensive X-ray characterization of the VFXB IGR J17062-6143 with \textit{NuSTAR}, \textit{XMM-Newton}, and \textit{Swift}. High resolution X-ray spectroscopy of the RGS spectra reveals evidence for an oxygen-rich absorbing medium and shows hints for a mildly-relativistic, shocked outflow. Secondly, broadband spectral fitting suggests the presence of a truncated accretion disk with an inner radius of $79^{+22}_{-18}$ $R_g$. Finally, an extensive pulsation search in the EPIC-pn data does not detect the recently reported pulations at 163 Hz using RXTE data \citep{Strohmayer17}.

\subsection{The nature of very-faint X-ray binaries}
\label{sec:magnetictruncation}

First, we will discuss two proposed explanations for the (sustained) very-faint nature of persistent VFXBs -- an ultra-compact orbit and magnetic inhibition -- and whether these can account for J1706's properties. We will also briefly discuss the possibility of both occuring in the same source. 

\subsubsection{Ultra-compact X-ray binary with a white dwarf donor?}
\label{sec:ucxb}

A possible origin for the low luminosities of VFXBs is the presence of an ultra-compact orbit \citep[][]{king06, intzand07, hameury16}. Such an UCXB might not be able to physically fit a large enough disk to sustain a higher accretion rate. In addition to a small orbital period, such systems could show a lack of H$\alpha$ emission from the disk as a hydrogen-rich donor does not fit in the small orbit \citep[e.g.][]{nelemans04, intzand09}. However, H$\alpha$ emission has been observed previously in a VFXB, ruling out a compact orbit as a universal explanation \citep{degenaar10}. Furthermore, a non-detection of H$\alpha$ does not necessarily imply an ultra-compact nature of the binary. Here, we will discuss whether the X-ray properties of J1706 contain hints towards an UCXB nature, focussing on the RGS continuum. 

The enhanced oxygen abundance measured in the RGS spectrum is unlikely to be of interstellar origin for two reasons: the interstellar oxygen abundance towards LMXBs is measured to be at most $\sim 1.3$ times the Solar abundance \citep{pinto13}. Secondly, we do not detect a similarly enhanced abundance of neon, which would expected if the excess oxygen was of interstellar nature. Instead, the neon edge is in fact well modelled by the $N_H$ value determined with lower-resolution instruments. Thus, the high oxygen abundance is more likely instrinsic to the source: possibly, outflowing material rich in oxygen could create a local overdensity of circumbinary absorbing material. In this scenario, such outflowing material is also expected to show blueshifted oxygen emission. Indeed the relatively broad emission feature around $18$ \AA~might be a combination of O VIII and O VII emission blueshifted by $\sim 0.05c$. 

In the above scenario, the accreted and expelled material is rich in oxygen. Hence, the donor star in this system might be a white dwarf (WD), requiring an ultra-compact orbit to allow Roche-lobe overflow. Identifying the nature of this possible WD is difficult using only the RGS spectra: on the one hand, the potential strong Ne X emission feature in the outflowing material might suggest that the donor is a O-Ne-Mg WD. On the other hand, as mentioned above, the neon edge is correctly modelled by merely the interstellar absorption. Alternatively, the donor could be a CO WD. However, the C-edge lies outside the RGS band, preventing us from directly investigating the C abundance in the RGS spectra. If the donor is indeed a CO WD, the strong Ne X emission line might simply result from the collisional ionisation, which tends to produce stronger Fe and Ne lines than O lines. The class of UCXBs contains several similar examples of possible CO or O-Ne-Mg WD donor identifications through X-ray spectroscopy: most prominently, HETG spectroscopy suggests the presence of such a WD donor in 4U 1626-67 \citep{schulz01b,krauss07}, while similar arguments have been made for several other (candidate) UCXBs \citep[see e.g.][]{juett01,juett03,juett05,nelemans04,nelemans06}. 

Recently, \citet{koliopanos13} and \citet{koliopanos14} suggested that the Fe K$\alpha$ line might be heavily surpressed in UCXBs with WD donors, as photons around $\sim 7$ keV would be mainly absorbed by oxygen instead of iron. J1706 shows a strong iron line in both \textit{XMM-Newton} and \textit{NuSTAR}, which is at odds with this statement if the donor is indeed a WD. However, \citet{madej14}, adjusting the \textsc{xillver} reflection model to CO WD donors, did not find an attenuation of the iron line. According to \citet{madej14}, this difference originates from solving the ionisation structure of the illuminated disk, instead of assuming a neutral gas as in \citet{koliopanos13}. 

A WD donor in J1706 is also consistent with the results from the recent extensive near-infrared, optical and UV investigation by {\color{blue} Hern\'andez Santisteban et al. (2017)}. Although the orbital parameters of J1706 could not be determined exactly, this SED analysis places an upper limit on the orbital period of $\sim$ one hour. Furthermore, they report a distinct lack of H$\alpha$ emission in a single-epoch optical spectrum, as expected in UCXBs \citep{nelson86,nelemans04,nelemans06,werner06}. The small orbit is required for a WD to Roche-lobe overflow such that we observe the system as an LMXB, while the lack of detected hydrogen is consistent with a WD donor. However, we reiterate that a lack of H$\alpha$ does not necessarily imply a compact orbit -- several LMXBs have been observed both with and without H$\alpha$ emission between different epochs (see {\color{blue} Hern\'andez Santisteban et al. (2017)} for a more detailed discussion) -- and a VFXB with $H\alpha$ emission has been observed as well \citep{degenaar10}. We also note that several UCXB are not very-faint X-ray binaries \citep{heinke13}. So merely being an UCXB is not sufficient to be a VFXB \'and VFXBs can not form a single subset of a larger class of UCXBs. 

\subsubsection{Magnetic inhibition}

An alternative proposed idea for the nature of persistently very faint accreting neutron stars, is that of magnetic inhibition of the accretion flow \citep{heinke15, degenaar14, degenaar17}. In this scenario, the neutron star magnetic field is strong enough to truncate the accretion disk away from the ISCO and as such prevent efficient accretion. In such a geometry, the magnetic field might also launch a propellor-driven outflow \citep[e.g.][]{illarionov75}. Through X-ray reflection spectroscopy, the inner disk radius can be measured to search for such a disk truncation. However, disk truncation is not direct evidence for magnetic inhibition, especially at low accretion rates where the accretion flow changes structure: the inner accretion disk can transition into a RIAF, also effectively resulting in a truncation of the thin disk \citep[e.g. ][]{narayan94}. 

\begin{figure*}
  \begin{center}
    \includegraphics[width=\textwidth]{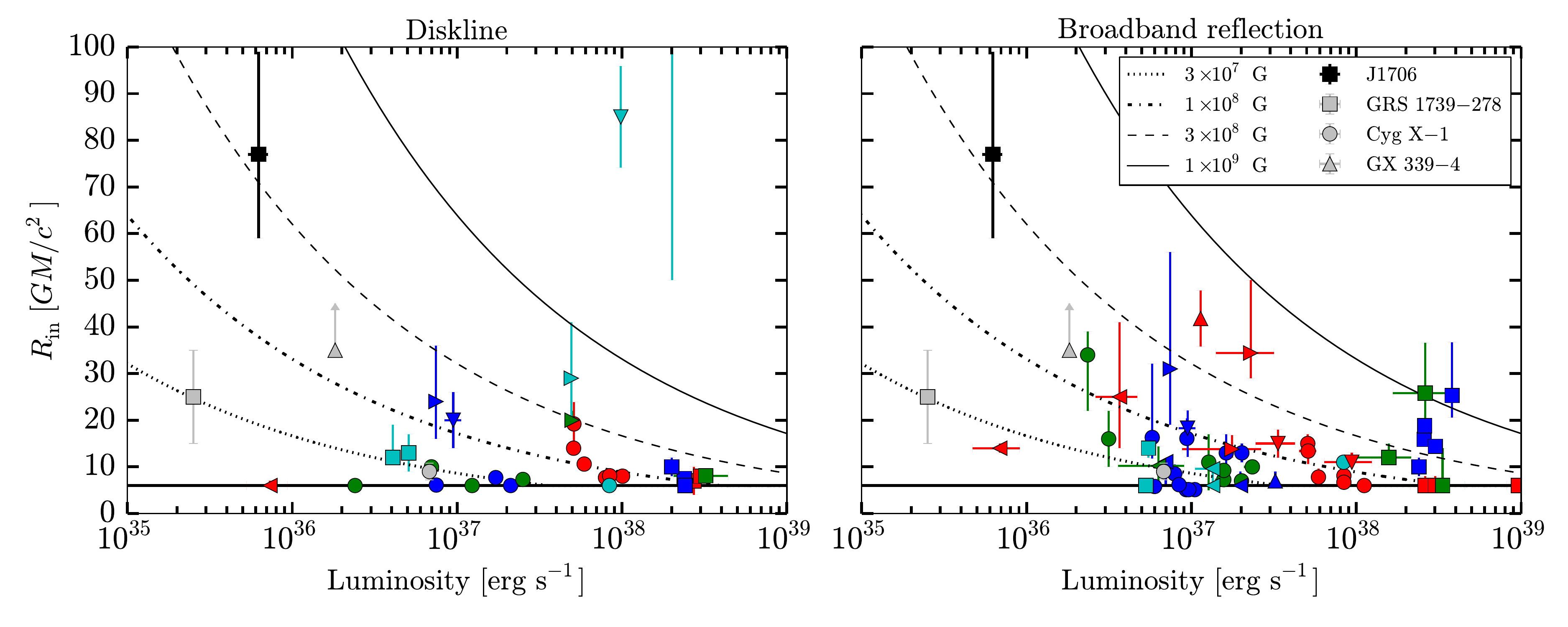}
    \caption{Overview of inner disk radius measurements as a function of bolometric X-ray luminosity ($0.3$--$79$ keV). The left panel shows estimates with the \textsc{diskline} model, while the right panel contains broadband reflection estimates (with e.g. \textsc{relxill} or \textsc{reflionx}). Each source has its own marker, listed in Table \ref{tab:sample} in Appendix \ref{sec:sample}. The thick black line indicates the ISCO, assuming the Schwarzschild metric, while the other black lines show curves of constant magnetic field assuming a magnetically truncated disk. \source~ and the black holes are shown in both panels for easy comparison. In grey, we plot three examples of BH LMXBs. We stress that this plot only presents global trends, and cannot be used for detailed inferences due to differences in the individual analyses. For details on the sources, references, and caveats see Appendix \ref{sec:sample}.}
    \label{fig:rin}
  \end{center}
\end{figure*}

Distinguishing between the formation of such a RIAF and magnetic truncation is problematic in a single source, but the comparison with the full sample of measured inner disk radii could help break the degeneracy. Hence, we present such a comparison in Figure \ref{fig:rin}: both panels show the measured inner disk radii versus X-ray luminosity of both a large sample of NSs and three BHs -- the latter, containing 2 BH LMXBs and the BH HMXB Cyg X-1, are selected for their coverage of the low X-ray luminosity regime (see Appendix \ref{sec:sample}). In the left panel, we plot estimates of the inner disk radius measured using the \textsc{diskline}-model, while in the right panel we plot those detemined using broadband reflection models such as \textsc{relxill} or \textsc{reflionx}. We also include our inner disk radius measurement for J1706 in both panels.

Appendix \ref{sec:sample} contains detailed information on source selection and the conversion to the used energy band ($3$--$79$ keV). In both panels, the black (dotted) lines indicate the predicted relation between inner radius and luminosity for a given magnetic field strength, following equation 1 in \citet{cackett09} and assuming magnetic disk truncation\footnote{We use standard geometrical and efficiency assumptions, and set $M=1.4M_{\odot}$ and $R=10$ km.}. We stress that, since the datapoints come from a heterogenous set of analyses and publications with different underlying assumptions and caveats, these plots only present global trends and cannot be used for any detailed inferences. For important caveats and differences between the publications and assumptions, we refer the reader to Appendex \ref{sec:sample} as well. 

Due to observational challenges, the low X-ray luminosity region of interest is highly undersampled, both in BHs and NSs: in addition to J1706, only a single inner radius measurement in a NS LMXB has been made below $L_X = 10^{36}$ erg s$^{-1}$ \citep{cackett10}. Prior analysis of high-quality \textit{XMM-Newton} spectra in three persistent VFXBs at even lower X-ray lumunosities \citep{armas13b} or two transient VFXBs \citep{armas11,armas13a} did not reveal any reflection features. Hence, the current data is not yet sufficient to distinguish between ADAF formation and disk truncation by the magnetic field at low luminosities. In addition, other sources show a truncated disk at much higher luminosities, where ADAF formation is less likely (for instance the Rapid Burster and the Bursting Pulsar), without being VFXBs; while a different type of disk-magnetic field interaction might be at play in those sources \citep[such as a trapped disk; ][]{dangelo10,degenaar14,vandeneijnden17} and the Bursting Pulsar has a very wide ($\sim 11.8$ day) orbit \citep{finger96}, this shows that magnetic truncation is not always sufficient to inhibit accretion and create a VFXB. 

So can magnetic truncation explain the persistent VFXB nature of J1706? Our measured inner disk radius of $77^{+22}_{-18}$ $R_g$ (assuming an inclination of $65^{\rm o}$) would be consistent with such a scenario. J1706 also shows a disk truncation significantly larger than typically observed in the complete sample. However, we cannot currently unambigiously infer the truncation's origin from either the source itself or the complete sample. There is also the additional complication of the unconstrained inclination: a low inclination (e.g. $25^{\rm o}$) yields a disk extending to the ISCO and cannot be excluded at $3\sigma$. If J1706 is indeed viewed at such a low inclination, magnetic truncation cannot account for its VFXB nature. While in such a scenario we might also expect to observe stronger reflection features (such as the currently undetected Compton hump) and a disk component (which we do not observe; see Section \ref{sec:bb}), we cannot exclude this possibility. 

We can however still estimate the magnetic field strength required to create the measured disk truncation; for this exercise, we apply equation 1 from \citet{cackett09}, assuming a mass of $1.4$ $M_{\odot}$, a radius of $10$ km, an accretion efficiency of $20\%$, and an anisotropy correction factor and a geometry conversion factor of unity. We convert the observed flux of $0.98\times10^{-10}$ erg s$^{-1}$ cm$^{-2}$ to luminosity using a distance of $7.3$ kpc \citep{keek17}. As this flux was calculated over the $0.3$--$79$ keV range, we do not apply an additional bolometric correction. The estimated field strength amounts to $B = (2.5 \pm 1.1) \times 10^8$ G. Based on the detected $163$ Hz pulsations, \citet{Strohmayer17} infer a similar magnetic field strength of $B \leq 3.5 \times 10^8$ G. If instead the source's inclination is low, and the disk truncates at the ISCO, we can put an $3\sigma$ upper limit on the magnetic field strength of $B \leq 4.8 \times 10^6$ G, lower than in any other accreting millisecond pulsar \citep{mukherjee15}. 

\subsubsection{Magnetic truncation in an UCXB?}

Finally, we briefly discuss the option of both magnetic inhibition and an ultra-compact orbit combined resulting in a VFXB. Such a scenario would not be particularly surprising: firstly, given the lower luminosities during outbursts of UCXBs, a smaller magnetic field is required to create a significant truncation. This is clearly shown by comparing \source~with the Bursting Pulsar \citep{kouveliotou96} in Figure \ref{fig:rin} (light blue downward triangle): while the measured inner radii of the two sources are the same within their uncertainties, the magnetic field required to cause the measured truncation is orders of magnitude higher in the Bursting Pulsar. Secondly, \citet{hameury16} suggest that in UCXBs the same instability might be responsible for the X-ray outbursts as in wider LMXBs, but occuring at much lower luminosities. This assumes a slightly different viscosity parameter in UCXBs, possibly due to a different composition of the accreted material. In this scenario, at their low-luminosities, UCXBs are thus not expected to behave the same as wider-orbit binaries, possibly still showing a thin disk. In other words, in this model UCXBs do not necessarily form the RIAF-geometry expected at low luminosity in wider LMXBs. Finally, the lack of hydrogen, resulting in a higher average number of nucleons per electron than in hydrogen-rich material, might impede efficient channeling of gas by the magnetic field lines, resulting in a more effective inhibition of the accretion flow. 

Could both mechanisms be at play in J1706 simultaneously? The recent SED investigation by {\color{blue} Hern\'andez Santisteban et al. (2017)} and the RGS continuum both suggest a UCXB nature of J1706, while the reflection modelling is consistent with magnetic truncation of the disk. If the companion in J1706 is indeed a WD, this might lead to the scenario as proposed by \citet{hameury16}. Moreover, J1706 would not be unique in this regard: a handful of UCXBs with WD donors \citep[XTE J1751-305, XTE J0929-314, XTE J1807-294, Swift J1756.9-2508 and NGC6440 X-2; e.g. ][]{patruno12} could possibly have a truncated disk as pulsations and channeled accretion are observed. However, also the combination of both mechanisms is not always sufficient: 4U 1626-67 is both an UCXB and possesses a strong magnetic field \citep{chakrabarty97}, but is not a VFXB. Hence, even in combination with magnetic inhibition, an ultra-compact orbit does not necessarily imply a low X-ray luminosity. 

\subsubsection{The origin of the blackbody emission}
\label{sec:bb}

As Type-I bursts are observed in J1706, part of the material must be accreted onto the NS surface. The SED modelling in J1706 by {\color{blue} Hern\'andez Santisteban et al. (2017)} shows that the X-ray blackbody component can not originate from the compact accretion disk. Together with the analysis in \citetalias{degenaar17}, we now have estimates of the X-ray blackbody parameters in three different epochs: 2014 (\textit{Chandra}), 2015 (\textit{Swift}) and 2016 (\textit{XMM-Newton}). Updating the \citetalias{degenaar17} results to a source distance of $7.3$ kpc, the blackbody radius is measured to be $5.9 \pm 0.2$ km, $8.1 \pm 0.9$ km and $7.2 \pm 0.2$ km in these three epochs respectively. While only the latter two are consistent, all three are larger than the expected size of a hotspot at the magnetic poles. Instead, they might originate from the NS surface itself, with its expected radius of $\sim 11-12$ km. Note that in our analysis, we find a lower blackbody radius ($4.8 \pm 1.3$ km) in the \textit{Swift} data than \citetalias{degenaar17}. This probably follows from the inclusion of a low-energy Gaussian to account for the $\sim 1$ keV excess in the EPIC-pn spectrum. 

The blackbody temperature has decreased from $0.48\pm0.01$ keV (2014) and $0.47\pm0.01$ keV (2015) in the first two epochs to $0.36 \pm 0.01$ keV (2016) in the final epoch. These values are consistent with the expected range of NS surface temperatures due to accretion predicted by \citet{zampieri95}, who estimate a temperature of $0.35$ keV at an accretion rate of $10^{-4}$ $L_{\rm Edd}$, increasing to $0.53$ keV at $10^{-3}$ $L_{\rm Edd}$. Indeed, the luminosity of J1706 was approximately two times higher in 2014 than in 2016, possibly accounting for part of the temperature change. Again, the addition of a $\sim 1$ keV Gaussian feature in the modelling of the 2016 data might also influence the measured temperature. We do note however that the calculations by \citet{zampieri95} do not include channeled accretion, which might be present in J1706 given the detection of pulsations in the \textit{RXTE} data. 

\subsection{High-resolution spectroscopy}

\subsubsection{A propellor-driven outflow?}

Here, we briefly discuss the potential detection of an outflow in the RGS spectra, before discussing a number of important caveats to the data and the analysis in section \ref{sec:caveats}. The EPIC-pn and RGS spectra of \source~indicate the presence of excess emission around an energy of $\sim 1$ keV, and an initial line search reveals hints of a number of narrow lines consistent with earlier \textit{Chandra} observations. A detailed analysis of the high-resolution RGS spectra reveals that this excess can be modelled as a blueshifted, collisionally ionised combination of both string narrow lines and a quasi-continuum of weaker lines. The blueshift of the linemodel ($z=-0.048$), which is consistent with those for the lines suggested by the linesearch in both RGS and HETG ($z=-0.03$ to $-0.6$), indicates a possible outflow. 

Our results are consistent with the analysis by \citetalias{degenaar17} of \textit{Chandra}-HETG spectra of J1706, taking into account the differences between the instruments. Our analysis of the RGS spectra has also allowed us to estimate the ionisation state of the outflow and yielded a more confident identification of the lines seen in both instruments. \citetalias{degenaar17} briefly discuss the option of an outflow with a blueshift of $z=-0.045$, which is consistent with our analysis, although the ionisation type could not be determined with \textit{Chandra} due to limited statistics. Two interesting possibilities for the interpretation of the potential outflow are a jet and, as suggested by \citetalias{degenaar17}, a propeller-driven wind. In the latter scenario, the magnetic field truncates the accretion disk (far) outside the co-rotation radius and creates a propeller expelling part of the matter from the accretion disk \citep{illarionov75}.

Detailed simulations of this propeller-regime by \citet{romanova09} revealed that such systems could simultaneously exhibit a jet and an conical wind, both powered by the NS magnetic field. For a standard NS mass and radius, \citet{romanova09} report typical jet velocities of $0.4$--$0.5$ c and typical wind velocities of $0.03$--$0.1$ c. The outflow detected in J1706 has an observed velocity of $0.048$ c, clearly most consistent with the conical wind. Even assuming an inclination of the source of $\sim 65^{\rm o}$, the observed velocity corresponds to merely $\sim 0.12$ c for a jet perpendicular to the accretion disk. A wind outflow is indeed what was suggested by \citetalias{degenaar17} as well based on the \textit{Chandra} spectroscopy of J1706. Part of the gas should of course still be accreted since Type-I bursts and pulsations have been observed in J1706. 

However, the outflow can only be adequately modelled as a collisionally ionised plasma (\textsc{bapec} in \textsc{xspec}), while a photo-ionised plasma does not describe the observed features. The former is typically associated with jets, while the latter is expected for a wind outflow. Hence, we might instead observe material from the outer regions of a jet that are less collimated and slower, but are still collisionally ionised. Another option might be that the outflowing material is collisionally ionised in shocks as the accretion flow interacts with the magnetospehere.

\subsubsection{Limitations to the analysis}
\label{sec:caveats}

It is important to discuss here a number of caveats in the presented analysis of the RGS spectrum and the interpretation of our results as an outflow. As stated before, our initial phenomenological line search uses two single-trial significances to find indications for narrow lines. Estimating the actual number of trials is problematic due to the different continuums, line widths, and the correlation between trials at nearby wavelengths. While the comparison of our results with the similar line search in \textit{Chandra} reduces the chance of false positives, caution should still be exercised when identifying individual lines. Hence, we use the line search primarily as a starting point for the subsequent line modeling.

However, this line modeling does not come without caveats either; firstly, at several wavelengths the first and second order RGS spectra show significant discrepensies. While the largest differences (for instance around $8$ and $14.5$ \AA) occur outside the the range that drives the line modelling ($\sim 10$ -- $13$ \AA), these differences do exceed the optimal systematics for the RGS detectors as reported by \citet{kaastra16}. Additionally, the strongest narrow lines in the \textsc{bapec}-model appear to miss the largest deviations from the continuum by a single to a few bins (see Figure \ref{fig:linemodel}). This might be partially attributed to the assumed Solar abundances in the line model, but the fit appears to be driven primarily by the pseudo-continuum of weaker lines fitting the $\sim 1$ keV excess. 

Given these three caveats, we cannot unambigously infer the presence of an outlow and we stress that caution must be exercised in interpreting the RGS spectra. However, we also note that out of all options we attempted -- a simple Gaussian, reflection of a normal or CO-rich accretion disk, a photo-ionised outflow and a collisionally-ionised outflow -- only the latter is even remotely capable of accounting for the broad $\sim 1$ keV excess. Moreover, the outflow suggested by the line modeling fits into the broader picture of J1706, connecting the possible magnetic propellor to the observed overabundance of oxygen in circumbinary material. Such an outflow is also suggested by the comparison between the mass tranfer rate and the mass accretion rate inferred from the X-ray luminosity of the system ({\color{blue} Hern\'andez Santisteban et al., 2017}). Finally, the full line-modelling and all identifications based on the line search independently point towards similar blueshifts in the emission.

\subsection{A lack of pulsations: is \source~an intermittent AMXP?}

Despite applying three different techniques, we do not detect the $163$ Hz pulsations recently claimed in the only \textit{RXTE} observation of J1706 \citep{Strohmayer17} in the \textit{XMM-Newton} observation. Instead, we find an upper limit on the pulsed fraction of $5.4$\%. There are several possible explanations for the lack of detected pulsations: firstly, given the low flux, we might simply not be sensitive enough. Indeed, pulsations have been detected in other sources below our upper limits; for instance, in HETE J1900.1-2455 pulsations have been detected at pulsed fractions down to $\sim 0.1\%$ \citep{patruno12c}. While we can exclude pulsations with a similar pulsed fraction as seen in \textit{RXTE} ($9.4\pm1.1\%$), the fraction might simply have dropped below our detection limit.\footnote{In the abstract, \citet{Strohmayer17} report instead a pulsed fraction of $5.54 \pm 0.67\%$, without mentioning it again. In the \textit{RXTE} data, we measure a fraction of $9.9\pm1.7$\%, and hence we adopt the $9.4\pm1.1\%$ value.}

Alternatively, J1706 might be an intermittent AMXP. These sources, of which three are currently known (SAX J1748-2021 \citep{altamirano08}, Aql X-1 \citep{casella08} and HETE J1900.1-2455 \citep{galloway07}, only show detectable pulsations a fraction of the time -- most extremely, Aql X-1 only shows the pulsations in a single $\sim 150$ s interval among $\sim 1.5$ Ms of \textit{RXTE} observations. In HETE J1900.1-2455, the pulsations disappeared around two months into the outbursts, only to sporadically reappear afterwards \citep{patruno12c} before the source returned to quiescence over a decade later \citep{degenaar17b}. \citet{patruno12c} discussed that the disappearance of the pulsations followed from the burial of the source's magnetic field by the accretion process, as proposed by \citet{cumming01}. A similar scenario could be at play in J1706, where the reported pulsations occured in $2008$, about a year into the outburst. As our observations where taken approximately $8$ years after the outburst start, the magnetic field might have gotten buried as well. However, we do not have the archival data required to test such an idea, and the pulsations in HETE J1900.1-2455 disappeared before the outburst was one year old (e.g. when J1706's pulsations were detected). 

The final possibility is of course that the reported \textit{RXTE} signal is a false positive; while significant, the signal is detected at a much lower significance ($4.3\sigma$ after excluding frequencies between $2048$ and $4096$ Hz in the search) than for instance the pulsations in Aql X-1 \citep[$\sim 9\sigma$,][]{casella08}. However, with a single \textit{RXTE} observation, there is no way to test this possibility. To further investigate the AMXP nature and properties of J1706, se will search for the pulsations again in approved future \textit{AstroSAT} observations. 

\section*{Acknowledgements}

We thank the anonymous referee for insightful comments that improved the quality of this paper. We thank Javier Garcia for providing the \textsc{xillver}$_{\rm CO}$ model tables. JvdE acknowledges the hospitality of the Institute of Astronomy in Cambridge, where part of this research was carried out. JvdE, ND and JVHS are supported by a Vidi grant from the Netherlands Organization for Scientific Research (NWO) awarded to ND. ND also acknowledges support via a Marie Curie fellowship (FP-PEOPLE-2013-IEF-627148) from the European Commission. JVHS also acknowledges partial support from NewCompStar, COST Action MP1304. CP and ACF are supported by ERC Advanced Grant Feedback 340442. The research leading to these results has received funding from the European Union's Horizon 2020 Programme under the AHEAD project (grant agreement n. 654215). AP is supported by an NWO Vidi grant. DA acknowledges support from the Royal Society. RW is supported by an NWO Top grant, module 1. This work is based on data from the NuSTAR mission, a project led by California Institute of Technology, managed by the Jet Propulsion Laboratory, and funded by NASA. We acknowledge the use of the Swift public data archive. The semi-coherent pulsation search was performed on the ATLAS computer cluster of the Max-Planck-Institut f\"ur Gravitationsphysik.




\input{output.bbl}



\appendix

\section{$L_x$ versus $R_{\rm in}$}
\label{sec:sample}

In Table \ref{tab:sample}, we list all measured fluxes and inner radii used to produce Figure \ref{fig:rin}, together with other information such as source distances, type of spectral model, telescope and ObsId, and reference. A machine-readable version of this table is available in the online materials -- the caveats listed below apply there as well. 

We have performed an extensive literature search in order to form a respresentative sample of NS LMXBs. However, we do not claim to present a complete sample and there is a possibility that individual measurements are missing. The three BH LMXBs are selected to cover the low luminosity range in order to provide a meaningfull comparison with \source. We also required the BH LMXB spectra to be fitted with a free inner radius, as the inner radius is often fixed to the ISCO in order to estimate the spin of the BH. 

Since the inner radius measurements were performed by different authors, we list several important notes and caveats about the composition and interpretation of the table below:
\begin{enumerate}
    \item Fluxes in the table are quoted in the range reported in the references. While we do apply a bolometric correction when we convert fluxes to bolometric luminosities for the plot (see below), we do not apply it to the fluxes in the table; this correction is prone to assumptions about the underlying spectral shape, and quoting the uncorrected fluxes allows for the application of other corrections if required. We also note that the corrections are generally small and do not alter the trends in the resulting plot significantly, especially since the luminosity-axis is logarithmic. 
    \item For the bolometric correction, we defined three \textit{'standard'} spectral shapes for different luminosity ranges. These models and their parameters (listed below) are obtained from the detailed spectral analysis in \citet{cackett10}. We used these standard spectra to convert the fluxes to the luminosity in the $0.3$--$79$ keV range. The standard spectral shapes are
    \begin{itemize}
        \item Uncorrected luminosity $< 5\times10^{36}$ erg s$^{-1}$: \textsc{powerlaw} with $\Gamma = 2.1$ and $N_{\rm po} = 0.07$ photons cm$^{-2}$ s$^{-1}$ keV$^{-1}$ at 1 keV, based on the spectrum of HETE J1900.1-2455 at these luminosities. 
        \item Uncorrected luminosity $\geq 5\times10^{36}$ and $<5\times10^{37}$ erg$^{-1}$: \textsc{diskbb+bbody+powerlaw} with $kT_{\rm disk} = 0.9$ keV, $N_{\rm disk} = 100$, $kT_{\rm BB} = 2.0$ keV, $N_{\rm BB} = 1.2$, $\Gamma = 3.5$ and $N_{\rm po} = 0.5$ photons cm$^{-2}$ s$^{-1}$ keV$^{-1}$ at 1 keV, based on the spectrum of 4U 1636-53 at these luminosities.
        \item Uncorrected luminosity $\geq 5\times10^{37}$ erg$^{-1}$: \textsc{diskbb+bbody} with $kT_{\rm disk} = 1.0$ keV, $N_{\rm disk} = 180$, $kT_{\rm BB} = 1.8$ keV and $N_{\rm BB} = 3.8$, based on the spectrum of Ser X-1 at these luminosities.
    \end{itemize}
    \item Often, both a \textsc{diskline}-model and a broadband reflection model are listed for the same ObsId. In these instances, the authors attempted both models and we list results from both approaches. These are plotted separately in the two panels in Figure \ref{fig:rin}.
    \item When a '/' is present in an ObsId, the different ObsIds are either analysed together by the authors or together form a single continuous observation.
    \item Often, spectra from multiple X-ray observatories are analysed simultaneously to characterise the reflection spectrum. In those instances, we list the observatory most important for fitting the reflection spectrum.
    \item We use the same distances to convert flux to luminosity as the authors of the original spectral analyses. We refer the reader to the referenced papers for any discussion on those distances. 
    \item If no uncertainty is quoted on flux, inner radius or distance, this uncertainty was not mentioned in the original reference. For most distance estimates, this occurs when the source is located in a globular cluster and has an accurate, independent distance measure. If this occurs for an inner disk radius measurement, obtaining this value was typically not a major goal of the analysis. 
    \item If the inner radius is quoted in units of $R_{\rm ISCO}$ in the original reference, we convert to $R_g$ by multiplying by $R_{\rm ISCO} = 6$ $R_g$. In other words, we assume that the spin parameter $a$ is zero. A very limited amount of references leaves the spin free or at a fixed non-zero value, resulting in inner radii between $5$--$6$ $R_g$. This does not alter the trends in the relation between X-ray luminosity and measured inner radius.
    \item In references marked with an asterisk, the uncertainties are quoted at the $90\%$ confidence level. Otherwise, the errors are quoted at the $1\sigma$ level. 
\end{enumerate}

\begin{table*}
 \begin{center}
 \caption{\small{Sample of inner disk radius measurements from reflection in NS LMXBs. Distances are given in kpc, fluxes in $10^{-8}$ erg s$^{-1}$ cm$^{-2}$ and inner radii $R_{\rm in}$ in $R_g$. Errors on flux and inner radius are quoted at the $1\sigma$ confidence level, unless the reference contains *, indicating a $90\%$ confindence level. Abbreviated source names are: $^a$ SAX J1808.4-3658; $^b$ HETE J1900.1-2455; $^c$ 1RXS J180408.9-34205; $^d$ XTE J1709-267; $^e$ IGR J17480-2446; $^f$ SAX J1748.9-2021.}}
  \label{tab:sample}
   \begin{tabular}{llllllll}
  \multicolumn{8}{c}{\textbf{Neutron star LMXBs}} \\
  \hline \hline
  Source & Distance & Observatory & ObsId & Model type & Flux & $R_{\rm in}$ & Reference \\ \hline \hline
Ser X-1 {\color{red} $\medblackcircle$} & $8.4$ & Suzaku & 401048010 & \textsc{Diskline} & $ 1.19 \pm 0.01 $ & $ 8.0 \pm 0.3 $ & \citet{cackett10}\\
 & & XMM & 0084020501 & \textsc{Diskline} & $ 0.59 \pm 0.01 $ & $ 14 \pm 1 $ & \citet{cackett10}\\
 & & Suzaku & 401048010 & \textsc{BBREFL} & $ 1.32 \pm 0.08 $ & $ 6 \pm 1 $ & \citet{cackett10}\\
 & & XMM & 0084020501 & \textsc{BBREFL} & $ 0.60 \pm 0.01 $ & $ 15 \pm 2 $ & \citet{cackett10}\\
 & & NuSTAR & 30001013002/4 & \textsc{Kerrdisk} & $ 0.61 $ & $ 10.6 \pm 0.6 $ & \citet{miller13}\\
 & & NuSTAR & 30001013002/4 & \textsc{Reflionx} & $ 0.61 $ & $ 7.8 \pm 1.8 $ & \citet{miller13}\\
 & & Chandra & 16208/9 & \textsc{Diskline} & $ 0.938 \pm 0.01 $ & $ 7.7 \pm 0.1 $ & \citet{chiang16a}* \\
 & & Chandra & 16208/9 & \textsc{Relflionx} & $ 0.938 \pm 0.01 $ & $ 7.1^{+1.1}_{-0.6} $ & \citet{chiang16a}* \\
 & & Suzaku & 408033010/20/30 & \textsc{Diskline} & $ 0.994 \pm 0.01 $ & $ 8.1^{+0.4}_{-1.2} $ & \citet{chiang16b}* \\
 & & Suzaku & 408033010/20/30 & \textsc{Relline} & $ 0.994 \pm 0.01 $ & $ 8.1 \pm 0.5 $ & \citet{chiang16b}* \\
 & & Suzaku & 408033010/20/30 & \textsc{BBREFL} & $ 0.994 \pm 0.01 $ & $ 6.7^{+0.8}_{-0.7} $ & \citet{chiang16b}* \\
 & & NuSTAR & 30001013002/4 & \textsc{Diskline} & $ 0.527 \pm 0.002 $ & $ 19.2 \pm 4.7 $ & \citet{matranga17b}* \\
 & & NuSTAR & 30001013002/4 & \textsc{rfxconv} & $ 0.527 \pm 0.062 $ & $ 13.4 \pm 2.8 $ & \citet{matranga17b}* \\
4U 1636-53 {\color{blue} $\medblackcircle$} & $6.0 \pm 0.1$ & XMM & 0303250201 & \textsc{Diskline} & $ 0.17 \pm 0.01 $ & $ 6.1^{+0.4}_{-0.1} $ & \citet{cackett10}\\
 & & XMM & 0500350301 & \textsc{Diskline} & $ 0.39 \pm 0.01 $ & $ 7.7 \pm 0.6 $ & \citet{cackett10}\\
 & & XMM & 0500350401 & \textsc{Diskline} & $ 0.48 \pm 0.01 $ & $ 6.0^{+0.2}_{-0.} $ & \citet{cackett10}\\
 & & XMM & 0303250201 & \textsc{BBREFL} & $ 0.18 \pm 0.02 $ & $ 8.5 \pm 2 $ & \citet{cackett10}\\
 & & XMM & 0500350301 & \textsc{BBREFL} & $ 0.37 \pm 0.01 $ & $ 13 \pm 4 $ & \citet{cackett10}\\
 & & XMM & 0500350401 & \textsc{BBREFL} & $ 0.46 \pm 0.03 $ & $ 13 \pm 2 $ & \citet{cackett10}\\
 & & NuSTAR & 30102014002 & \textsc{Kyrline} & $ 0.212 \pm 0.001 $ & $ 5.12 \pm 0.15 $ & \citet{wang17}\\
 & & NuSTAR & 30102014004 & \textsc{Kyrline} & $ 0.112 \pm 0.001 $ & $ 5.12 \pm 0.10 $ & \citet{wang17}\\
 & & NuSTAR & 30102014006 & \textsc{Kyrline} & $ 0.116 \pm 0.001 $ & $ 5.12 \pm 0.08 $ & \citet{wang17}\\
 & & NuSTAR & 30102014002 & \textsc{Relxill} & $ 0.121 \pm 0.001 $ & $ 5.8 \pm 0.7 $ & \citet{wang17}\\
 & & NuSTAR & 30102014004 & \textsc{Relxill} & $ 0.113 \pm 0.001 $ & $ 16.1^{+4.3}_{-1.4} $ & \citet{wang17}\\
 & & NuSTAR & 30102014006 & \textsc{Relxill} & $ 0.117 \pm 0.001 $ & $ 16.3^{+15.8}_{-4.5} $ & \citet{wang17}\\
 & & NuSTAR & 30101024002 & \textsc{Relxill} & $ 0.089 \pm 0.002 $ & $ 6.18 \pm 0.18 $ & \citet{ludlam17a}\\
4U 1705-44 {\color{green} $\medblackcircle$} & $5.8 \pm 0.2$ & Suzaku & 401046010 & \textsc{Diskline} & $ 0.17 \pm 0.01 $ & $ 10.0 $ & \citet{cackett10}\\
 & & Suzaku & 401046020 & \textsc{Diskline} & $ 0.61 \pm 0.01 $ & $ 7.3 \pm 0.4 $ & \citet{cackett10}\\
 & & Suzaku & 401046030 & \textsc{Diskline} & $ 0.30 \pm 0.01 $ & $ 6^{+0.2} $ & \citet{cackett10}\\
 & & XMM & 0402300201 & \textsc{Diskline} & $ 0.043 \pm 0.001 $ & $ 6^{+1} $ & \citet{cackett10}\\
 & & Suzaku & 401046010 & \textsc{BBREFL} & $ 0.57 \pm 0.05 $ & $ 10 $ & \citet{cackett10}\\
 & & Suzaku & 401046020 & \textsc{BBREFL} & $ 0.49 \pm 0.01 $ & $ 7 \pm 1 $ & \citet{cackett10}\\
 & & Suzaku & 401046030 & \textsc{BBREFL} & $ 0.31 \pm 0.01 $ & $ 11 \pm 6 $ & \citet{cackett10}\\
 & & XMM & 0402300201 & \textsc{BBREFL} & $ 0.042 \pm 0.001 $ & $ 34^{+5}_{-12} $ & \citet{cackett10}\\
 & & Suzaku & 406076010 & \textsc{Reflionx} & $ 0.0334 \pm 0.0003 $ & $ 16 \pm 6 $ & \citet{disalvo15}* \\
 & & NuSTAR & 30101025002 & \textsc{Reflionx} & $ 0.34 \pm 0.03 $ & $ 7.26 \pm 1.08 $ & \citet{ludlam17a}\\
 & & NuSTAR & 30101025002 & \textsc{BBREFL} & $ 0.34 \pm 0.02 $ & $ 9.24 \pm 0.48 $ & \citet{ludlam17a}\\
4U 1820-30 {\color{cyan} $\medblackcircle$} & $7.6 \pm 0.4$ & Suzaku & 401047010 & \textsc{Diskline} & $ 1.21 \pm 0.01 $ & $ 6.0 \pm 0.1 $ & \citet{cackett10}\\
 & & Suzaku & 401047010 & \textsc{BBREFL} & $ 1.21 \pm 0.13 $ & $ 11 \pm 1 $ & \citet{cackett10}\\
GX 17+2 {\color{red} $\medblacksquare$} & $9.8 \pm 0.4$ & Suzaku & 402050010 & \textsc{Diskline} & $ 2.37 \pm 0.01 $ & $ 7 \pm 3 $ & \citet{cackett10}\\
 & & Suzaku & 402050010 & \textsc{Diskline} & $ 2.60 \pm 0.01 $ & $ 8 \pm 0.4 $ & \citet{cackett10}\\
 & & Suzaku & 402050020 & \textsc{BBREFL} & $ 2.27 \pm 0.01 $ & $ 6^{+2} $ & \citet{cackett10}\\
 & & Suzaku & 402050020 & \textsc{BBREFL} & $ 2.62 \pm 0.01 $ & $ 6^{+2} $ & \citet{cackett10}\\
 & & NuSTAR & 30101023002 & \textsc{Reflionx} & $ 7.3 \pm 0.9 $ & $ 6.06 \pm 0.06 $ & \citet{ludlam17a}\\
 & & NuSTAR & 30101023002 & \textsc{BBREFL} & $ 7.3 \pm 0.6 $ & $ 6^{+0.12} $ & \citet{ludlam17a}\\
 GX 349+2 {\color{blue} $\medblacksquare$} & $9.2$ & NuSTAR & 30201026002 & \textsc{Reflionx} & $ 2.224 \pm 0.001 $ & $ 15.9 \pm 1.5 $ & \citet{coughenour17}\\
 & & NuSTAR & 30201026002 & \textsc{Reflionx} & $ 2.239 \pm 0.001 $ & $ 18.8^{+1.6}_{-1.4} $ & \citet{coughenour17}\\
 & & NuSTAR & 30201026002 & \textsc{Reflionx} & $ 2.604 \pm 0.001 $ & $ 14.4 \pm 0.9 $ & \citet{coughenour17}\\
 & & NuSTAR & 30201026002 & \textsc{Reflionx} & $ 3.288 \pm 0.002 $ & $ 25.3^{+11.4}_{-4.7} $ & \citet{coughenour17}\\
 & & Suzaku & 400003010 & \textsc{Diskline} & $ 2.35 \pm 0.03 $ & $ 7.5 \pm 0.4 $ & \citet{cackett10}\\
 & & Suzaku & 400003020 & \textsc{Diskline} & $ 1.97 \pm 0.01 $ & $ 10 \pm 2 $ & \citet{cackett10}\\
 & & XMM & 0506110101 & \textsc{Diskline} & $ 2.38 \pm 0.01 $ & $ 6.0^{+0.5} $ & \citet{cackett10}\\
 & & Suzaku & 400003010 & \textsc{BBREFL} & $ 2.36 \pm 0.13 $ & $ 10 \pm 2 $ & \citet{cackett10}\\
Cyg X-2 {\color{green} $\medblacksquare$} & $11 \pm 2$ & Suzaku & 403063010 & \textsc{Diskline} & $ 2.22 \pm 0.01 $ & $ 8.1 \pm 0.9 $ & \citet{cackett10}\\
 & & Suzaku & 403063010 & \textsc{BBREFL} & $ 2.30 \pm 0.01 $ & $ 6^{+8} $ & \citet{cackett10}\\
 & & NuSTAR & 30001141002 & \textsc{Reflionx} & $ 1.58 \pm 0.01 $ & $ 25.8 \pm 10.8 $ & \citet{mondal17}* \\
 & & NuSTAR & 30001141002 & \textsc{Reflionx} & $ 0.95 \pm 0.01 $ & $ 12^{+3}_{-1.5} $ & \citet{mondal17}* \\
 \hline  
  \end{tabular}
  \end{center}
\end{table*}

\begin{table*}
 \begin{center}
  \contcaption{\small{}}
     \begin{tabular}{llllllll}
  \hline \hline
  Source & Distance & Observatory & ObsId & Model type & Flux & $R_{\rm in}$ & Reference \\ \hline \hline
SAX J1808$^a$ {\color{cyan} $\medblacksquare$} & $3.5 \pm 0.1$ & Suzaku & 903003010 & \textsc{Diskline} & $ 0.20 \pm 0.01 $ & $ 12^{+7}_{-1} $ & \citet{cackett10}\\
 & & XMM & 0560180601 & \textsc{Diskline} & $ 0.25 \pm 0.01 $ & $ 13 \pm 4 $ & \citet{cackett10}\\
 & & Suzaku & 903003010 & \textsc{BBREFL} & $ 0.27 \pm 0.02 $ & $ 14 \pm 2 $ & \citet{cackett10}\\
 & & XMM & 0560180601 & \textsc{BBREFL} & $ 0.26 \pm 0.01 $ & $ 6^{+0.2} $ & \citet{cackett10}\\
HETE J1900$^b$  {\color{red} $\medblacktriangleleft$} & $3.6 \pm 0.5$ & Suzaku & 402016010 & \textsc{Diskline} & $ 0.034 \pm 0.001 $ & $ 6^{+1} $ & \citet{cackett10}\\
 & & Suzaku & 402016010 & \textsc{BBREFL} & $ 0.032 \pm 0.005 $ & $ 14 \pm 1 $ & \citet{cackett10}\\
 & & XMM & 0671880101 & \textsc{Reflionx} & $ 0.136 \pm 0.008 $ & $ 25^{+16}_{-11} $ & \citet{papitto13}* \\
1RXS J1804$^c$  {\color{blue} $\medblacktriangleleft$} & $5.8$ & NuSTAR & 90101003002 & \textsc{Reflionx} & $ 0.48 \pm 0.001 $ & $ 6.0^{+3} $ & \citet{ludlam16}* \\
 & & NuSTAR & 80001040002 & \textsc{Relxill} & $ 0.171 $ & $ 11.1_{-5.1} $ & \citet{degenaar16b}\\
4U 1608-52  {\color{green} $\medblacktriangleleft$} & $3.7 \pm 0.8$ & NuSTAR & 90002002002 & \textsc{Relxillp} & $ 0.2 $ & $ 10.2 \pm 4.2 $ & \citet{degenaar15}* \\
4U 1728-34  {\color{cyan} $\medblacktriangleleft$} & $4.6 \pm 0.5$ & NuSTAR & 80001012002 & \textsc{Reflionx} & $ 0.6 $ & $ 6.0^{+4.8} $ & \citet{sleator16}* \\
 & & NuSTAR & 80001012002 & \textsc{Relxill} & $ 0.6 $ & $ 9.6 \pm 2.4 $ & \citet{sleator16}* \\
XTE J1709$^d$ {\color{red} $\medblacktriangleright$} & $8.5 \pm 0.1$ & NuSTAR & 90201025002/3 & \textsc{Reflionx} & $ 0.2 \pm 0.1 $ & $ 13.8^{+3}_{-1.8} $ & \citet{ludlam17b}* \\
 & & NuSTAR & 90201025002/3 & \textsc{Reflionx} & $ 0.26 \pm 0.1 $ & $ 34.4^{+15.6}_{-5.4} $ & \citet{ludlam17b}* \\
4U 1702-429 {\color{blue} $\medblacktriangleright$} & $4.19 \pm 0.15$ & XMM & 0604030101 & \textsc{diskline} & $ 0.4 $ & $ 24^{+12}_{-8} $ & \citet{iaria16}* \\
 & & XMM & 0604030101 & \textsc{rfxconv} & $ 0.4 $ & $ 31^{+25}_{-12} $ & \citet{iaria16}* \\
IGR J17480$^e$ {\color{green} $\medblacktriangleright$} & $5.5 \pm 0.0$ & Chandra & 13161 & \textsc{Diskline} & $ 1.01 \pm 0.05 $ & $ 20 \pm 2 $ & \citet{miller11}\\
SAX J1748$^f$ {\color{cyan} $\medblacktriangleright$} & $8.5$ & XMM & 0748391301 & \textsc{Diskline} & $ 0.55 $ & $ 29^{+12}_{-9} $ & \citet{pintore16}* \\
Aql X-1 {\color{red} $\medblacktriangledown$} & $5.2 \pm 0.7$ & NuSTAR & 80001034002/3 & \textsc{Reflionx} & $ 1.17 $ & $ 15 \pm 3 $ & \citet{king16}\\
EXO 1745-248 {\color{blue} $\medblacktriangledown$} & $5.5$ & XMM & 0744170201 & \textsc{Diskline} & $ 0.26 \pm 0.03 $ & $ 20 \pm 6 $ & \citet{matranga17a}* \\
 & & XMM & 0744170201 & \textsc{rdblur} & $ 0.26 \pm 0.03 $ & $ 18.3^{+3.8}_{-6.2} $ & \citet{matranga17a}* \\
J1706 {\color{black} $\medblacksquare$} & $7.3 \pm 0.5$ & NuSTAR & 30101034002 & \textsc{Diskline} & $ 0.0098 \pm 0.0001 $ & $ 77^{+22}_{-18}$ & -\\
GRO J1744-28 {\color{cyan} $\medblacktriangledown$} & $8.0$ & Chandra & 16605/6 & \textsc{Diskline} & $ 0.974 \pm 0.001 $ & $ 85.0 \pm 10.9 $ & \citet{degenaar14}\\
 & & NuSTAR & 80002017004 & \textsc{Diskline} & $ 2.62 \pm 0.02 $ & $ 130^{+240}_{-80} $ & \citet{younes15}\\
MXB 1730-335 {\color{red} $\medblacktriangleup$} & $7.9$ & NuSTAR & 90101009002 & \textsc{Reflionx} & $ 0.147 $ & $ 41.8 \pm 6 $ & \citet{vandeneijnden17}\\
GX 3+1 {\color{blue} $\medblacktriangleup$} & $6.1$ & XMM & 0655330201 & \textsc{rdblur} & $ 0.7 $ & $ 7^{+2}_{-1} $ & \citet{pintore15}* \\
\\
\multicolumn{8}{c}{\textbf{Accreting black holes}} \\
\hline \hline
  Source & Distance & Observatory & ObsId & Model type & Flux & $R_{\rm in}$ & Reference \\ \hline \hline
GRS 1739-278 {\color{Gray} $\medblacksquare$} & $8.5$ & NuSTAR & 80101050002 & \textsc{Relxill} & $ 0.00291 \pm 0.00006 $ & $ 25 \pm 10 $ & \citep{furst16}* \\
GX 339-4 {\color{Gray} $\medblacktriangleup$} & $8.0$ & Suzaku & 403067010 & \textsc{Laor} & $ 0.024 $ & $\geq35$ & \citep{tomsick09}* \\
Cyg X-1 {\color{Gray} $\medblackcircle$} & $1.86$ & NuSTAR & 30001011007 & \textsc{Relxillp} & $ 1.68 $ & $ 9 \pm 1.8 $ & \citep{parker15}\\
  \hline 
  \end{tabular}
  \end{center}
\end{table*}



\bsp	
\label{lastpage}
\end{document}